\documentclass{aastex}
\usepackage{spr-astr-addons}

\RequirePackage{color}

\begin{document}

\title{First Stars. I. Evolution without mass loss}

\slugcomment{}
%% Running heads
\shorttitle{First stars: evolution without mass loss}
\shortauthors{D. Bahena \& J. Klapp}

\author{D. Bahena\altaffilmark{1}} \and \author{J. Klapp\altaffilmark{2}}

\altaffiltext{1}{Astronomical Institute of the Academy of Sciences,\\
              Bo\v{c}n\'{\i} II 1401, 14131 Praha 4, Czech Republic.\\
              e-mail: bahen@universo.com}
\altaffiltext{2}{Instituto Nacional de Investigaciones Nucleares,\\
              Km 36.5 Carretera M\'exico-Toluca s/n, Salazar, Ocoyoacac, \\
              52750 Estado de M\'exico, Mexico.\\
              e-mail: jaime.klapp@inin.gob.mx}

\footnote{Accepted for publication in Astrophysics \& Space Science}

\begin{abstract}

The first generation of stars was formed from primordial gas.
Numerical simulations suggest that the \textit{first stars} were
predominantly very massive, with typical masses $M\geq 100
M_{\odot}$. These stars were responsible for the reionization of the
universe, the initial enrichment of the intergalactic medium with
heavy elements, and other cosmological consequences. In this work,
we study the structure of Zero Age Main Sequence stars for a wide
mass and metallicity range and the evolution of $100$, $150$, $200$,
$250$ and $300 M_{\odot}$ galactic and pregalactic Pop III very
massive stars without mass loss, with metallicity $Z=10^{-6}$ and
$10^{-9}$, respectively. Using a stellar evolution code, a system of
10 equations together with boundary conditions are solved
simultaneously. For the change of chemical composition, which
determines the evolution of a star, a diffusion treatment for
convection and semiconvection is used. A set of 30 nuclear reactions
are solved simultaneously with the stellar structure and evolution
equations. Several results on the main sequence, and during the
hydrogen and helium burning phases, are described. Low metallicity
massive stars are hotter and more compact and luminous than their
metal enriched counterparts. Due to their high temperatures,
pregalactic stars activate sooner the triple alpha reaction
self-producing their own heavy elements. Both galactic and
pregalactic stars are radiation pressure dominated and evolve below
the Eddington luminosity limit with short lifetimes. The physical
characteristics of the first stars have an important influence in
predictions of the ionizing photon yields from the first luminous
objects; also they develop large convective cores with important
helium core masses which are important for explosion calculations.
\end{abstract}


\section{Introduction}
\label{sec:introduction}

A first generation of stars composed of primordial nearly pure H/He
gas must have existed, since heavy elements can only be synthesized
in the interior of the stars. These \textit{first stars}, also
called Population III (or Pop III), were responsible for the initial
enrichment of the intergalactic medium (IGM) with heavy elements
\citep{Bond1981, Klapp1981, Klapp1983, Klapp1984, Cayrel1986,
Cayrel1996, Carr1987, Carr1994, Bromm2002}.

UV photons radiated by the \textit{first stars}, perhaps together
with an early population of quasars, are expected to have
contributed to the IGM reionization \citep{Haiman1997, Ferrara1998,
Miralda2000, Tumlinson2000, Bromm2001, Schaerer2002}. The energy
output from the \textit{first stars} might have left a measurable
imprint on the cosmic microwave background (CMB) on very small
scales \citep{Visniac1987, Dodelson1995}.

However, despite an intense observational effort, the discovery of a
true Pop III remains elusive. To probe the time when star formation
first started entails observing at very high redshifts $z\gtrsim
10$.

The \textit{first stars} were typically many times more massive and
luminous than the Sun \citep{Larson2004}. A review of theoretical
results on the formation of the \textit{first stars} has been made
by \citep{Bromm2004}. The masses of the first star-forming clumps
would have been about 500 to $1,000 M_{\odot}$. Several numerical
simulations suggest that the \textit{first stars} were predominantly
Very Massive Stars (VMS), with typical masses $M\gtrsim 100
M_{\odot}$ \citep{Bromm1999, Bromm2002, Nakamura2001, Abel2000,
Abel2002} and these stars had important effects on subsequent galaxy
formation.

In another scenario, the hypothesis that \textit{first stars} were
VMS ($M>140 M_{\odot}$) has been strongly criticized because the
pair-instability supernovae yield patterns are incompatible with the
Fe-peak and \textit{r}-process abundances found in very metal poor
stars. Models including Type II supernova and/or \textit{hypernova}
from zero-metallicity progenitors with $M=8-40 M_{\odot}$ can better
explain the observed trends \citep{Tumlinson2004}. The same authors
also pointed out that the sole generation of VMS ($M>140M_{\odot})$
cannot be possible and suggested that some VMS could be formed as
companions of stars with masses $M<140M_{\odot}$. Their Initial Mass
Function (IMF) proposition match quite well with the reionization
and nucleosynthesis evidence. Tegmark et al. (1997) argued that the
minimum baryonic mass is redshift dependent and lies in the range
$\sim10^{3.7}$ to $10^{6}M_{\odot}$, for $z \sim 10$ and $\sim 15$,
respectively, and that a participation of $\sim 10^{-3}$ of the
whole baryonic matter in the generation of luminous stars is
sufficient to reheating the universe by $z \sim 30$.

The Wilkinson Microwave Anisotropy Probe has observed the
large-angle polarization anisotropy of the CMB \citep{Cen2003,
Kogut2003, Sokasian2003, Wyithe2003}. Some measurements have been
interpreted as a signature of a substantial early activity of
massive star (MS) formation at high redshifts $z\gtrsim 15$.

The supernova explosions that ended the lives of the \textit{first
stars} were responsible for the initial enrichment of the
intergalactic medium with heavy elements \citep{Ostriker1996,
Gnedin1997, Bromm2003, Yoshida2004}. An interesting possibility
unique to zero-metallicity massive stars is the complete disruption
of their progenitors in pair-instability supernovae explosions,
which are predicted to leave no remnant behind \citep{Barkat1967,
Ober1983, Bond1984, Fryer2001, Heger2002, Heger2003, Bromm2004}. The
later works consider that this peculiar explosion mode could have
played an important role in quickly seeding the intergalactic medium
with the first metals.

Related to the \textit{first stars}, there are two very important
unsolved questions: 1. What are their typical masses and Initial
Mass Function (IMF)? and, 2. During their cuasi-static evolutionary
phases, do they have radiation driven winds or mass loss due to
other mechanisms?

 Star formation and accretion calculations suggest that the
\textit{first stars} were very massive \citep{Omukai2003}. On the
other hand, by comparing the observed abundance patterns of
Extremely Metal Poor (EMP) stars with supernova explosion
calculations, some authors have concluded that the \textit{first
stars} are more likely to have masses in the range $\sim 20-30
M_{\odot}$, but not more massive that $130 M_{\odot}$
\citep{Umeda2002, Heger2002}.

We have suggested that a possible solution to this inconsistency is
that \textit{first stars} are born very massive but that during
their cuasi-static evolutionary phases, lose mass and reach the
pre-supernova stage with the masses required from supernova
calculations to reproduce the EMP abundance pattern
\citep{Klapp2005, Bahena2006}. It is then very important to estimate
the amount of \textit{first stars} mass loss during the cuasi-static
evolutionary phases.

Kudritzki (2000, 2002) calculated wind models of massive stars
between 100 and $300 M_{\odot}$ and metallicities in the range
0.0001 to 1.0 solar, in an effective temperature range from $40,000$
to $60,000$ K, with the objective of predicting mass-loss rates at
very low metallicities applicable to the first generation of massive
stars. It was found that for very low metallicities, the line driven
mechanism becomes very inefficient and wind solutions can only be
obtained very close to the Eddington limit. He also pointed out that
very massive stars are pulsationally unstable, which might
contribute to stellar mass loss, in particular at low metallicity
when the contribution of the radiative driving to the wind
decreases. However, the critical mass for the onset of the nuclear
pulsational instability is uncertain.

Other mechanisms could induce \textit{first stars} mass loss, for
example, the low metallicity rotation models of Meynet and Maeder
(2002) show fast rotating cores that lose significant amounts of
mass and thus angular momentum. Rotation by enhancing the luminosity
and lowering the effective gravity increases the mass loss rate.
Then, it is possible that pulsation, rotation and Luminous Blue
Variables (LBV) type phenomenae could induce significant amounts of
mass loss during the \textit{first stars} cuasi-static evolutionary
phases.

Motivated by the above arguments, in a series of papers we will
study the structure, evolution and nucleosynthesis of the
\textit{first stars} with and without mass loss and rotation. In
this Paper I we present the evolutionary results without mass loss
and with no-rotation.

This work is organized as follows: In Sect. \ref{sec:modelling} we
describe the initial conditions of the stellar models and the way in
which the main physical variables are computed. Then, in Sect.
\ref{sec:results} we describe the main results, and in Sect.
\ref{sec:discussion} we discuss our results and compare them with
other authors. Finally, in Sect. \ref{sec:conclusions} we outline
our conclusions.

\section{Numerical modelling and input physics}
\label{sec:modelling}

We present Zero Age Main Sequence (ZAMS) models for stars in the
mass range $1-10,000 M_{\odot}$ with compositions
$(X,Z)=(0.765,10^{-2})$ for Pop I, $(X,Z)=(0.749,10^{-3})$ for Pop
II, $(X,Z)=(0.765,10^{-6})$ for galactic Pop III, and
$(X,Z)=(0.765,10^{-9})$ and $(1.0,10^{-10})$ for pregalactic Pop
III.

In this paper, evolutionary models for Pop III stars have been
calculated without mass loss and with no-rotation. The chemical
composition of the models is $(X,Z)=(0.765,10^{-6})$ and $(0.765,
10^{-9})$, for galactic and pregalactic stars, respectively. For the
evolution the chosen stellar masses were $100$, $150$, $200$, $250$
and $300 M_{\odot}$.

The main difference between galactic and pregalactic stars is their
initial metallicity. According to Castellani (2000) we consider a
Pop III metallicity range from $Z=10^{-6}$ to $10^{-10}$. The
transition from pregalactic to galactic stars occur at the critical
metallicity $Z\sim 10^{-3.5}Z_{\odot}$ \citep{Bromm2004}.
Pregalactic stars are characterized by self-producing their own
heavy elements; galactic stars correspond to the next generation of
stars which have been previously enriched with metals.

The computer program used for the calculations has been described by
Klapp (1981, 1983) and Bahena (2006, 2007), with updated input
physics.

For the nuclear reaction rates we use the Nuclear Astrophysics
Compilation of Reaction Rates (NACRE) by Angulo et al. (1999). A
diffusion treatment for convection and semiconvection is used. For
the opacity we have adopted the OPAL radiative opacities
\citep{RogersIglesias1992, IglesiasRogers1993}.

\section{Results}
\label{sec:results}

\subsection{Zero Age Main Sequence}

\begin{figure*}
\begin{center}
\includegraphics [width=154mm]{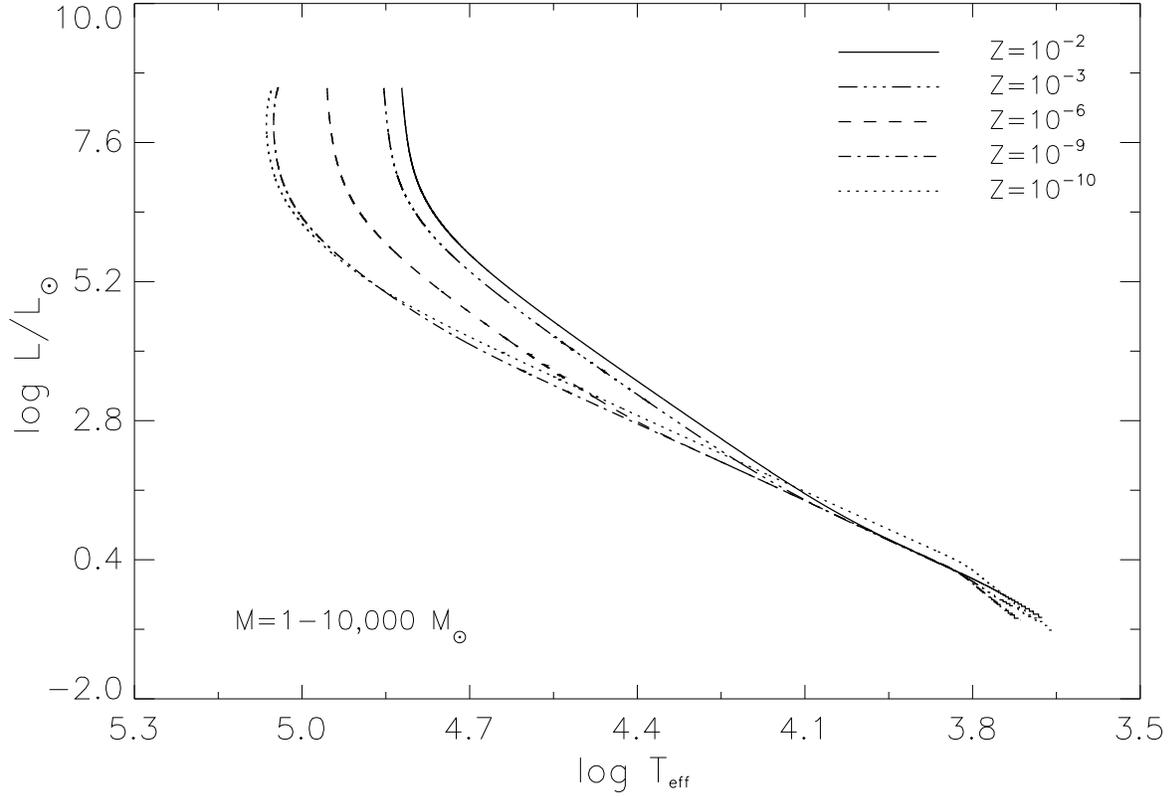}
%  \vspace*{174pt}
\caption{Hertzsprung-Russell (HR) diagram for ZAMS Pop I, II and III
stars for the mass range $1-10,000 M_\odot$ and metallicities
$Z=10^{-2}, 10^{-3}, 10^{-6}, 10^{-9}$ and $10^{-10}$.}
\label{bar_fig1}
\end{center}
\end{figure*}

For our Pop I, II and III models, in Fig. \ref{bar_fig1} we show
their ZAMS Hertzsprung-Russell (HR) diagram for the mass range
$1-10,000 M_{\odot}$. The objective is to understand the ZAMS
structure differences as function of mass and metallicity. A large
number of very detailed ZAMS models with different masses and
metallicities have been calculated.

The luminosity and effective temperature increases with mass. Low
mass stars are located at the right lower part of the diagram, while
massive and very massive stars are found in the left upper part,
because they are the most luminous and hotter. Pop III stars are
hotter than their Pop I and II counterparts, and so their locus on
the HR diagram is shifted to the left upper part. Pregalactic stars
are bluer than galactic stars. This is shown in Fig. 1 and also in
Fig. 2 that is an amplification for the $100-10000 M_{\odot}$ range.
In Fig. 3 we show a log $\varrho_c$ - log $T_c$ diagram for a large
mass and metallicity range.

\begin{figure}
\begin{center}
\includegraphics [width=84mm]{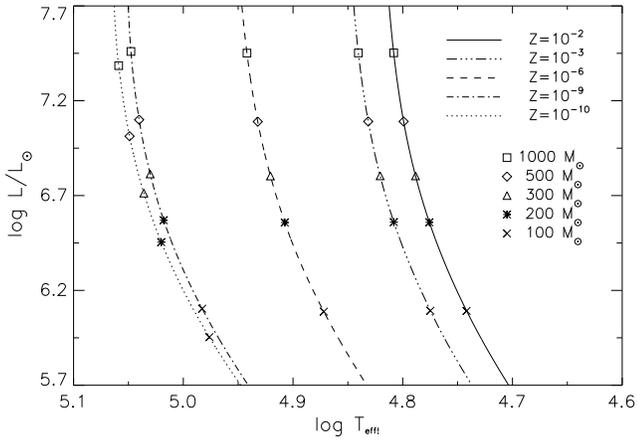}
%  \vspace*{174pt}
\caption{HR-diagram for ZAMS Pop I, II and III stars for the mass
range $100-10,000 M_{\odot}$. Depending on metallicity the Pop III
ZAMS is systematically shifted to higher effective temperature.}
\label{bar_fig2}
\end{center}
\end{figure}

\begin{figure}
\begin{center}
\includegraphics [width=84mm]{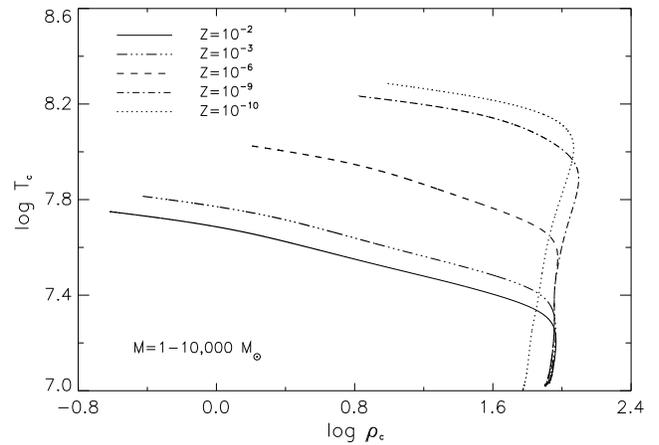}
%  \vspace*{174pt}
\caption{ZAMS Pop I, II, and III Log $\varrho_c$-Log $T_c$ plane for
the mass range $1-10,000 M_{\odot}$. With increasing density and
temperature, radiation pressure becomes more important.}
\label{bar_fig3}
\end{center}
\end{figure}

In Figs. 4 to 9 we show the main physical variables for the ZAMS
models. All quantities are plotted as function of the mass. As the
mass increases, ZAMS stars become bigger, brighter and less dense.
With decreasing metallicity, Pop III stars get very hot and compact.
All massive and very massive stars are dominated by radiation
pressure and develop a large convective core. In all cases, however,
their luminosity is below the Eddington upper luminosity limit.

\begin{figure}
\begin{center}
\includegraphics [width=84mm]{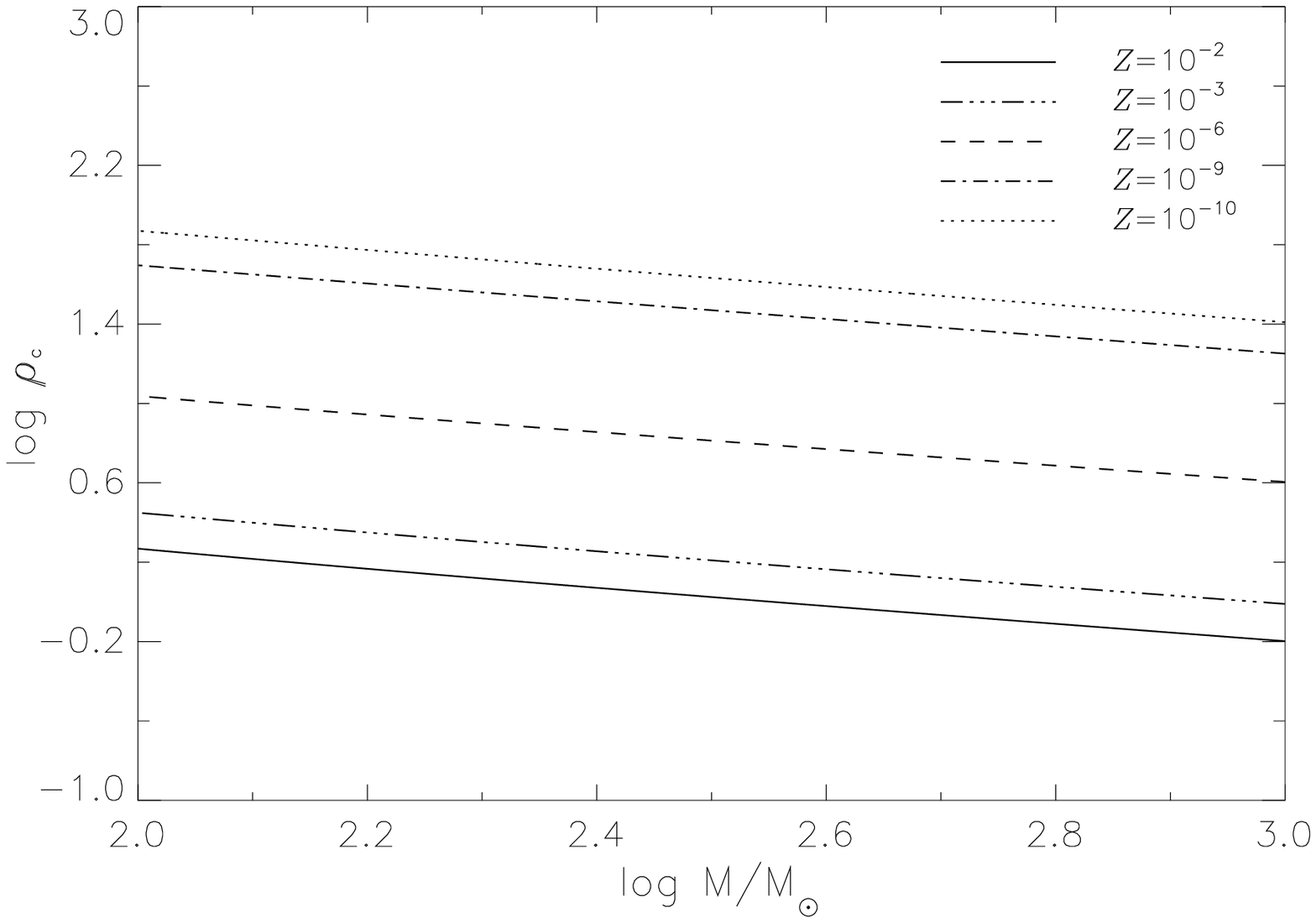}
%  \vspace*{174pt}
\caption{Central density for ZAMS models in the mass range $1-10,000
M_{\odot}$.} \label{bar_fig4}
\end{center}
\end{figure}

\begin{figure}
\begin{center}
\includegraphics [width=84mm]{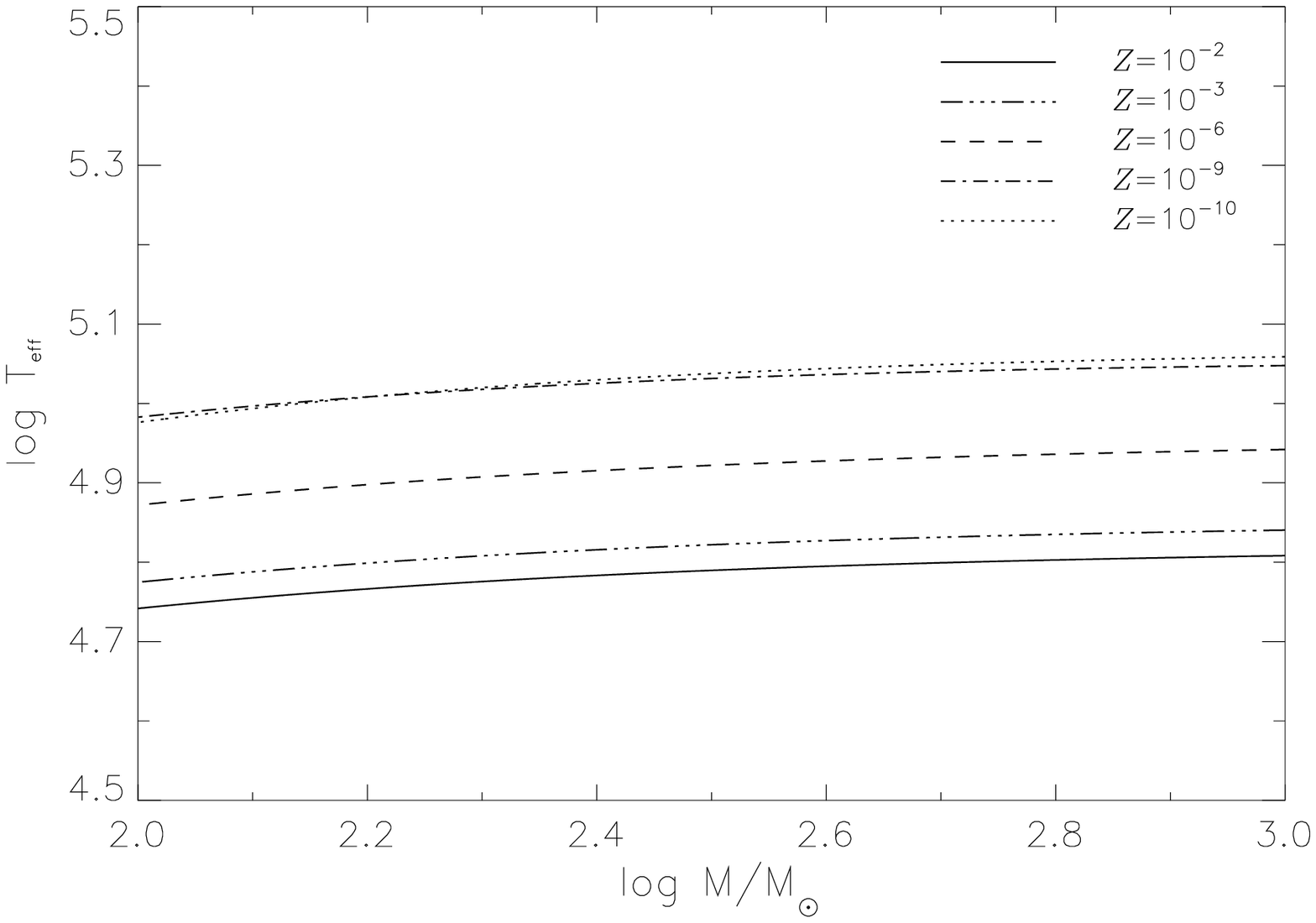}
%  \vspace*{174pt}
\caption{\emph{Ibidem}. Effective temperature.} \label{bar_fig5}
\end{center}
\end{figure}

\begin{figure}
\begin{center}
\includegraphics [width=84mm]{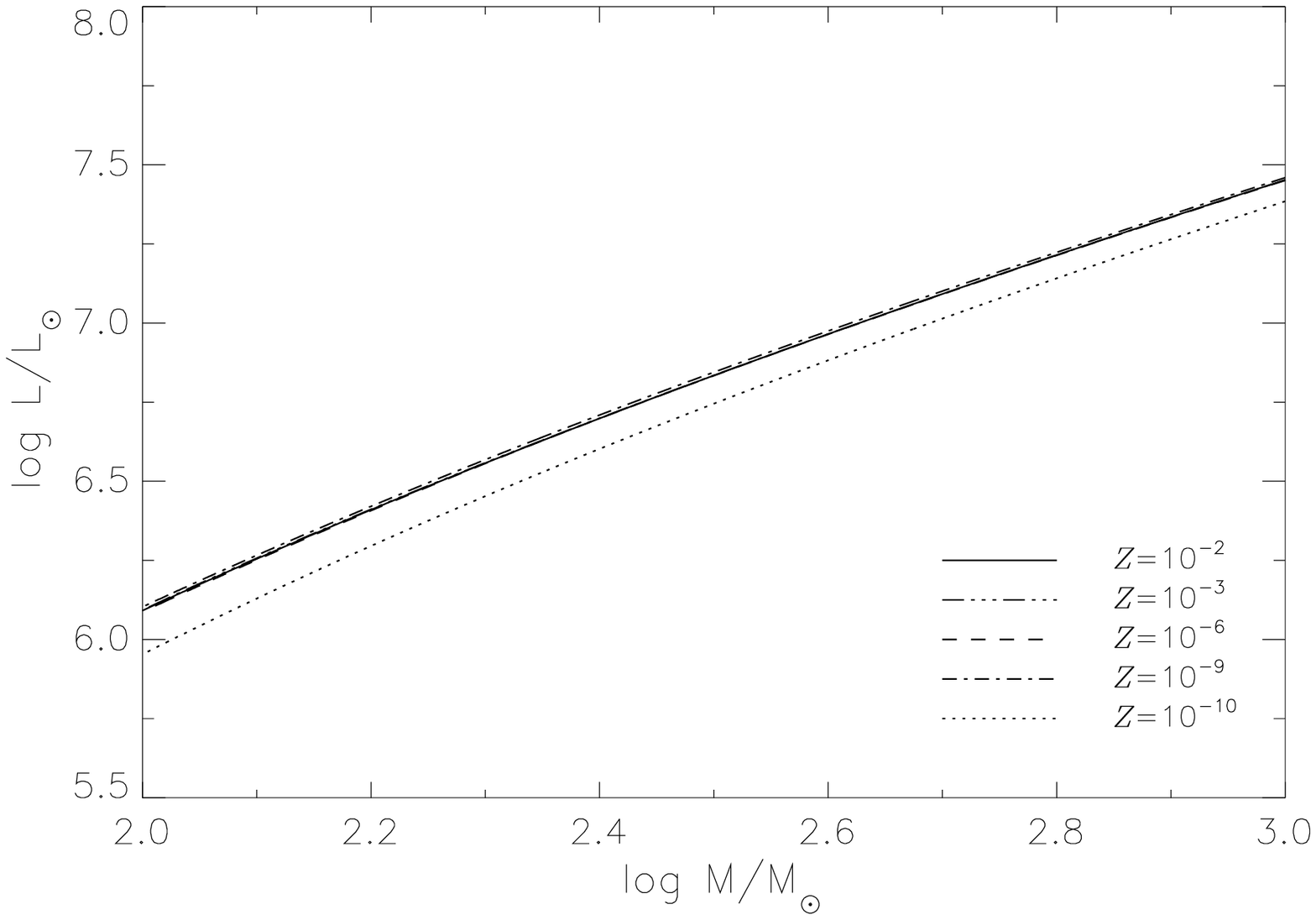}
%  \vspace*{174pt}
\caption{\emph{Ibidem}. Luminosity.} \label{bar_fig6}
\end{center}
\end{figure}

\begin{figure}
\begin{center}
\includegraphics [width=84mm]{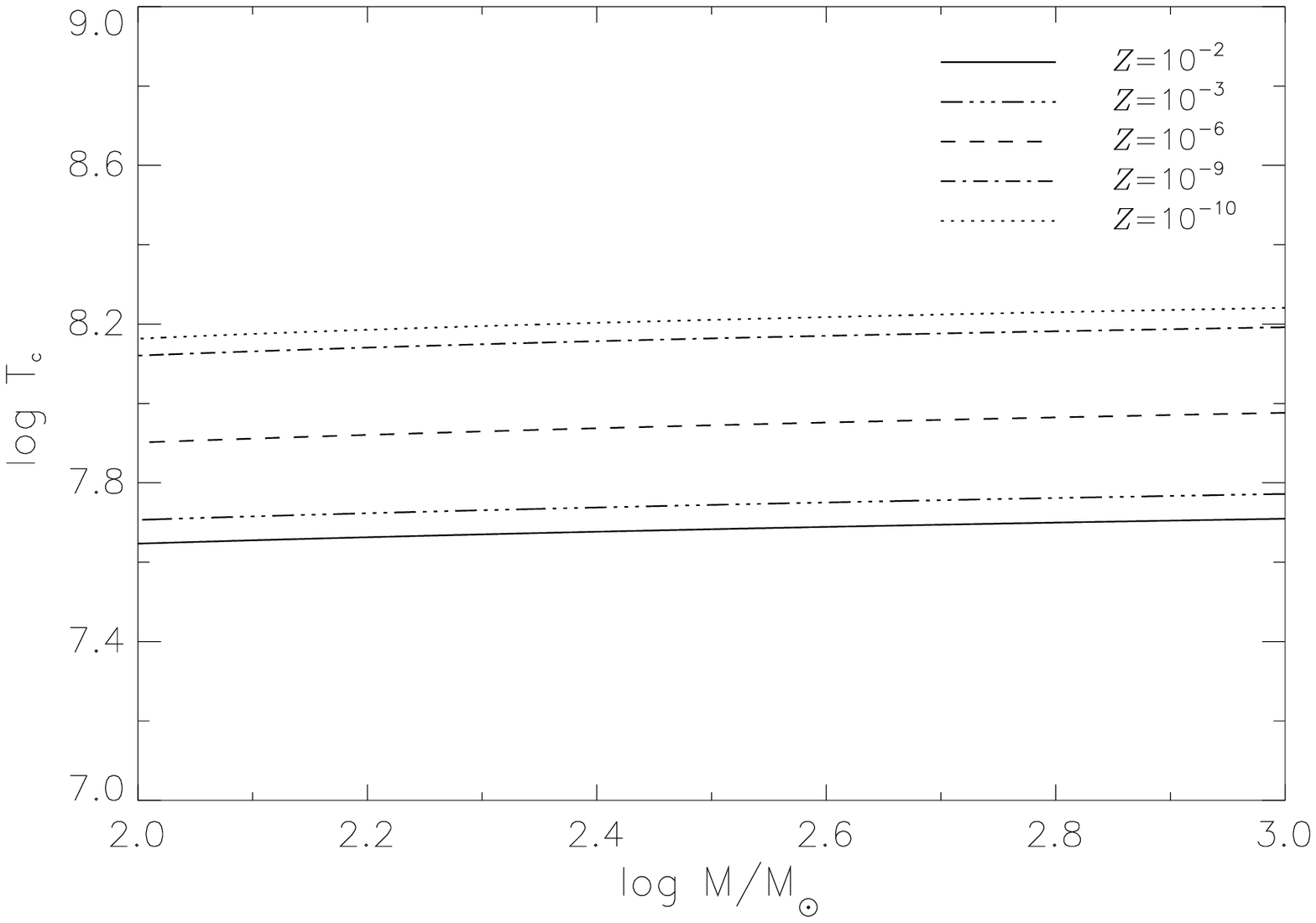}
%  \vspace*{174pt}
\caption{\emph{Ibidem}. Central temperature.} \label{bar_fig7}
\end{center}
\end{figure}

\begin{figure}
\begin{center}
\includegraphics [width=84mm]{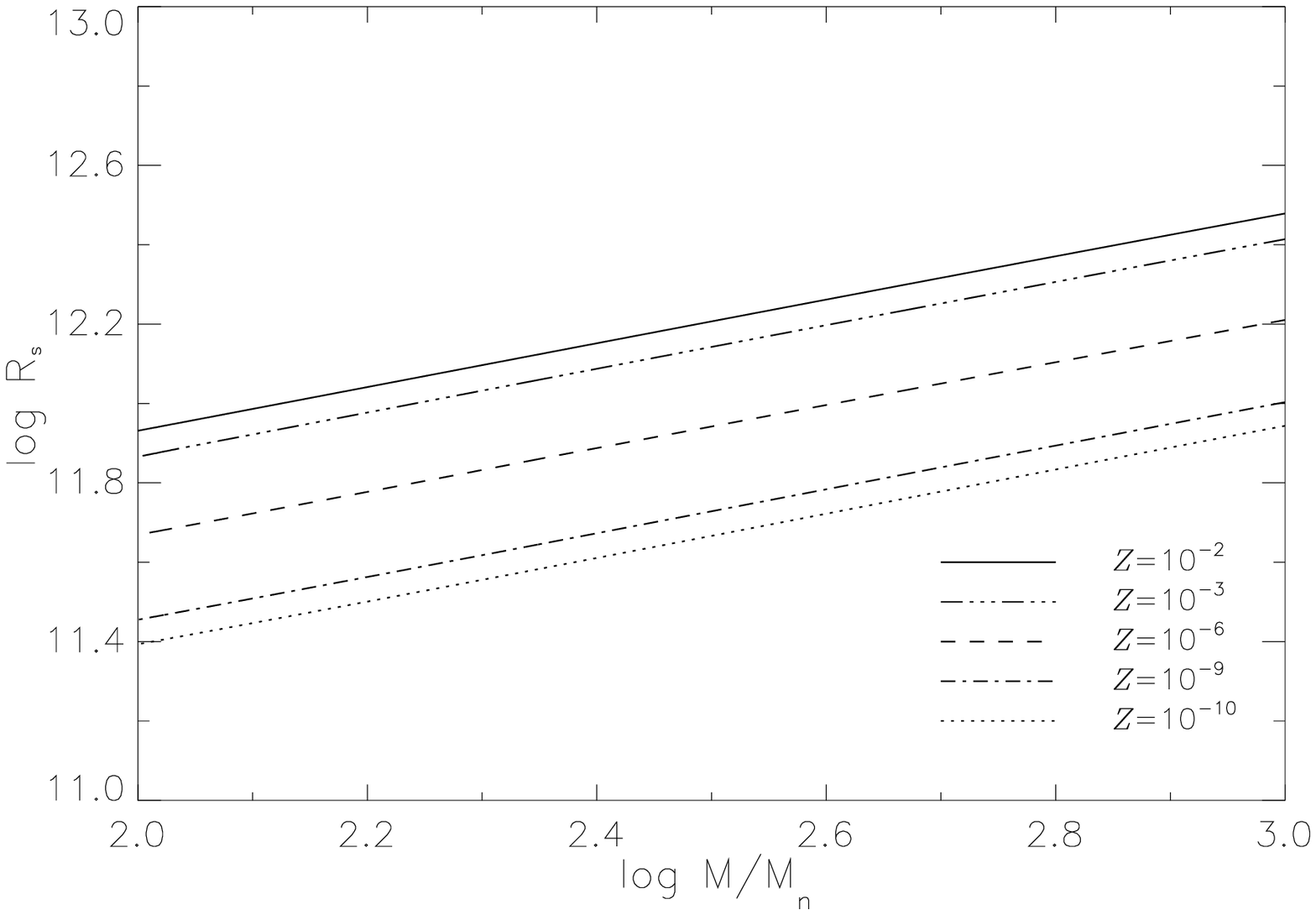}
%  \vspace*{174pt}
\caption{\emph{Ibidem}. Radius.} \label{bar_fig8}
\end{center}
\end{figure}

\begin{figure}
\begin{center}
\includegraphics [width=84mm]{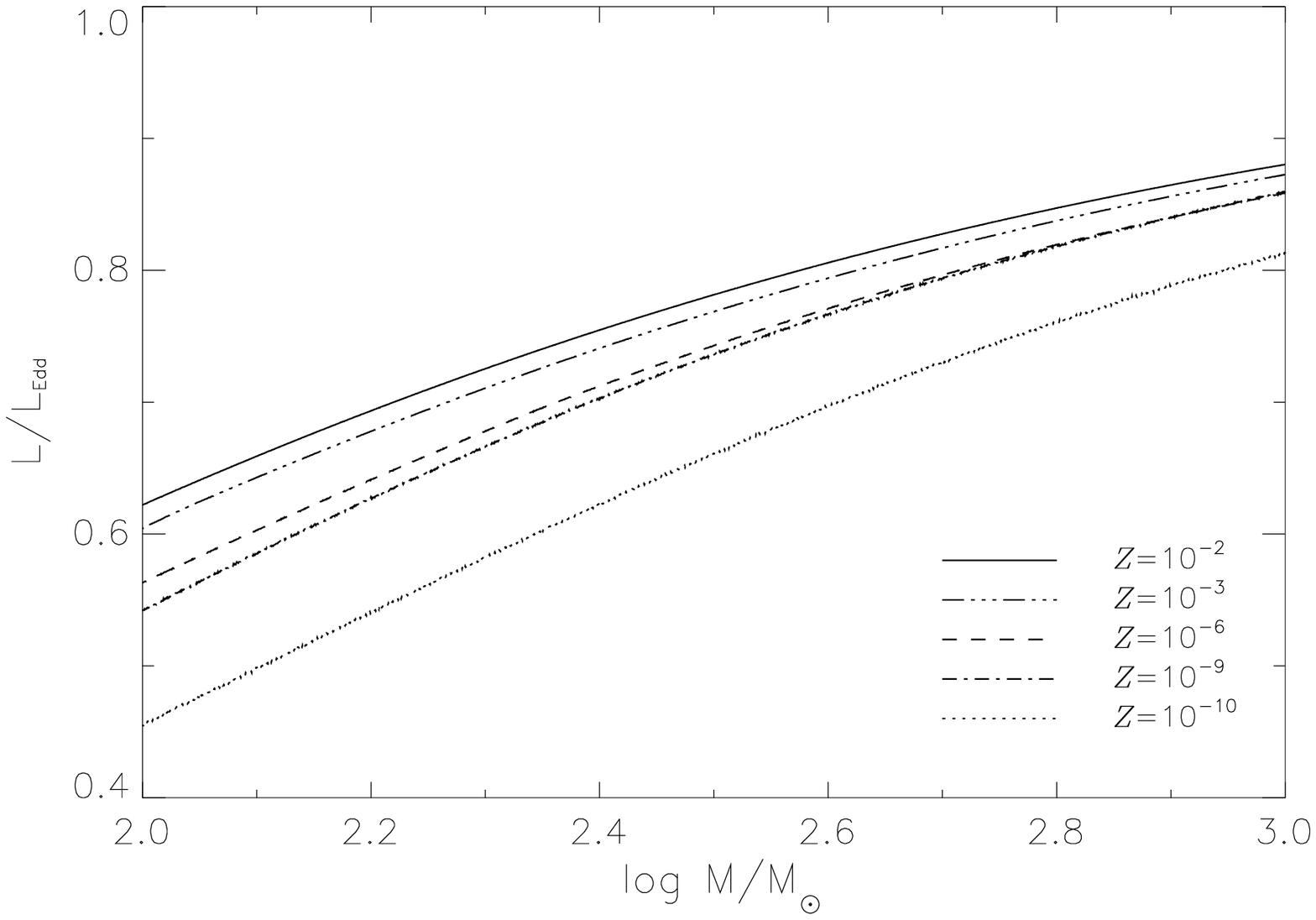}
%  \vspace*{174pt}
\caption{\emph{Ibidem}. Gamma factor which is the ratio of the
luminosity to the Eddington Luminosity.} \label{bar_fig9}
\end{center}
\end{figure}

Then, we present ZAMS models in the metallicity range from
$Z=10^{-10}$ to $10^{-6}$ for $M=100$, $250$, $500$, $750$ and $1000
M_{\odot}$, which are shown in Figs. 10 to 15. The central density
increases with decreasing metallicity. The most massive stars are
less dense. MS and VMS develop a large convective core, but its size
decreases slowly with decreasing metallicity. Stellar radius is
lower for low-metallicity. The radius also decrease with decreasing
mass, however, lower-metallicity stars are more compact. Central
temperature, and effective temperature, increase with decreasing
metallicity, and the most massive stars are hotter. Luminosity does
not depend on metallicity but on the mass. The most massive stars
are the most luminous stars. The Eddington luminosity factor
$\Gamma=L/L_{edd}$, which is the ratio of the luminosity $L$ to the
Eddington luminosity $L_{edd}$ does not depend on metallicity. The
most massive stars are the ones closest to the Eddington luminosity,
and the upper luminosity limit.

\begin{figure}
\begin{center}
\includegraphics [width=84mm]{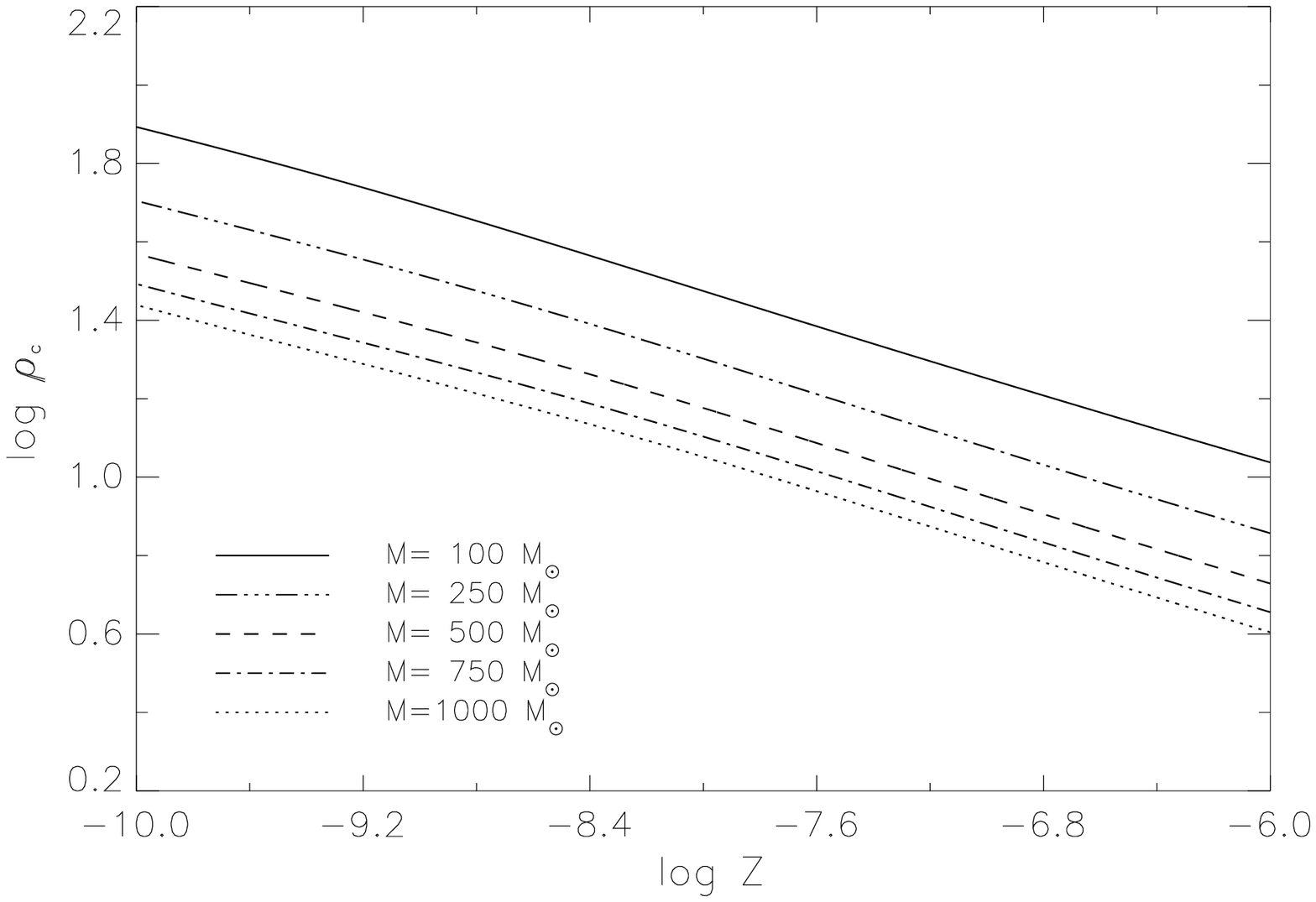}
%  \vspace*{174pt}
\caption{Central density in the metallicity range from $Z=10^{-10}$
to $10^{-6}$ for $M=100$, $250$, $500$, $750$ and $1000 M_{\odot}$.}
\label{bar_fig10}
\end{center}
\end{figure}

\begin{figure}
\begin{center}
\includegraphics [width=84mm]{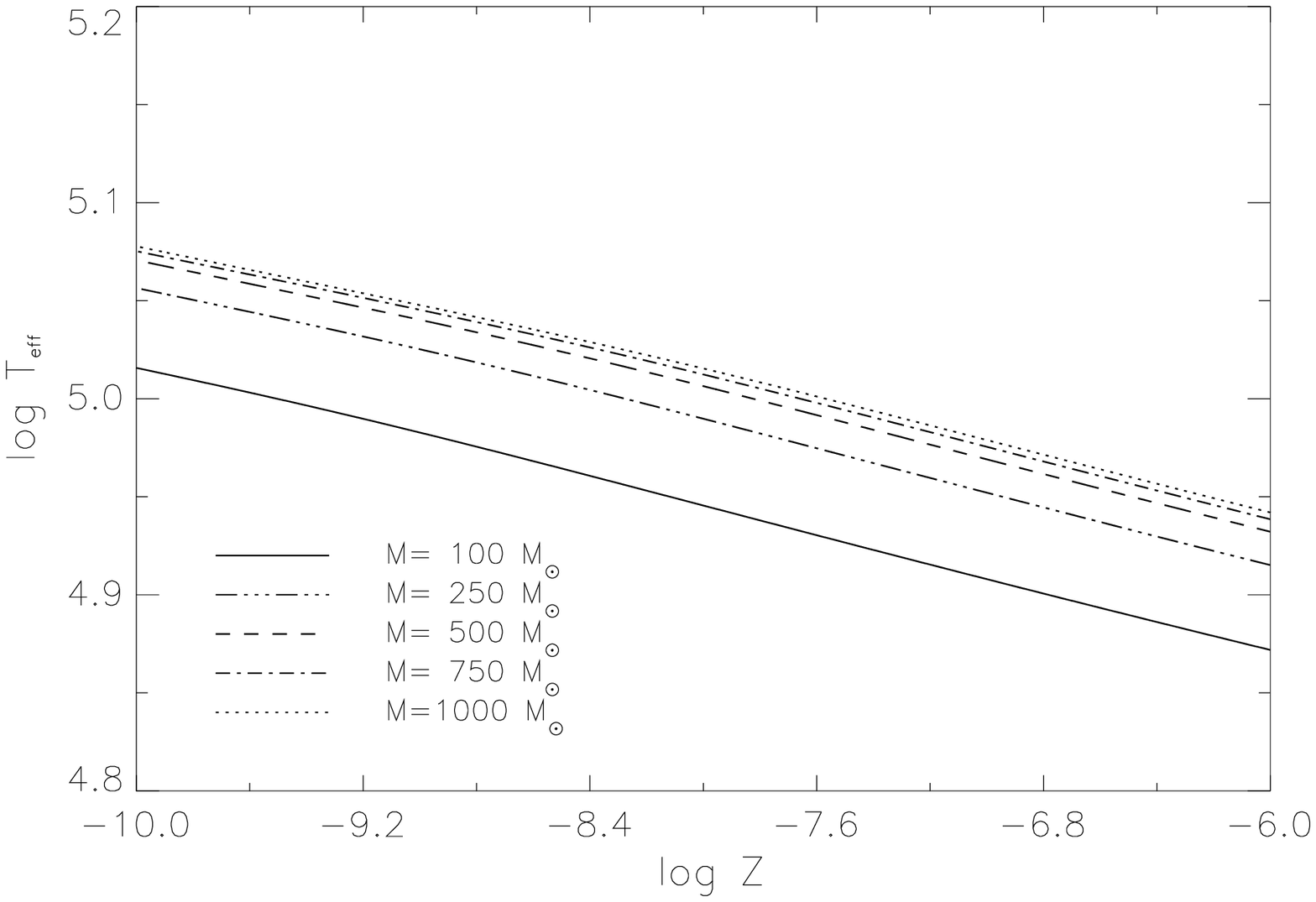}
%  \vspace*{174pt}
\caption{\emph{Ibidem}. Effective temperature.} \label{bar_fig11}
\end{center}
\end{figure}

\begin{figure}
\begin{center}
\includegraphics [width=84mm]{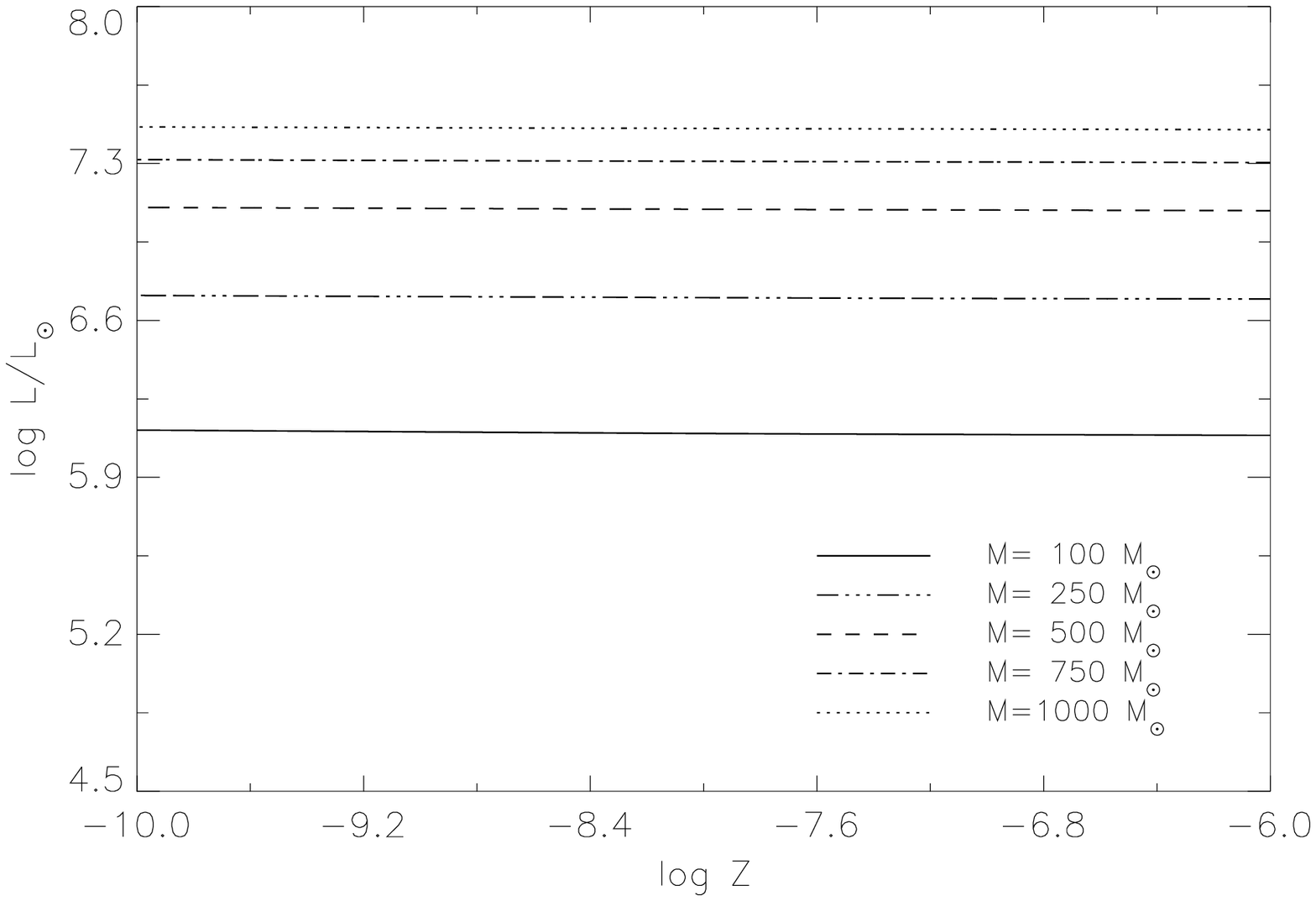}
%  \vspace*{174pt}
\caption{\emph{Ibidem}. Luminosity.} \label{bar_fig12}
\end{center}
\end{figure}

\begin{figure}
\begin{center}
\includegraphics [width=84mm]{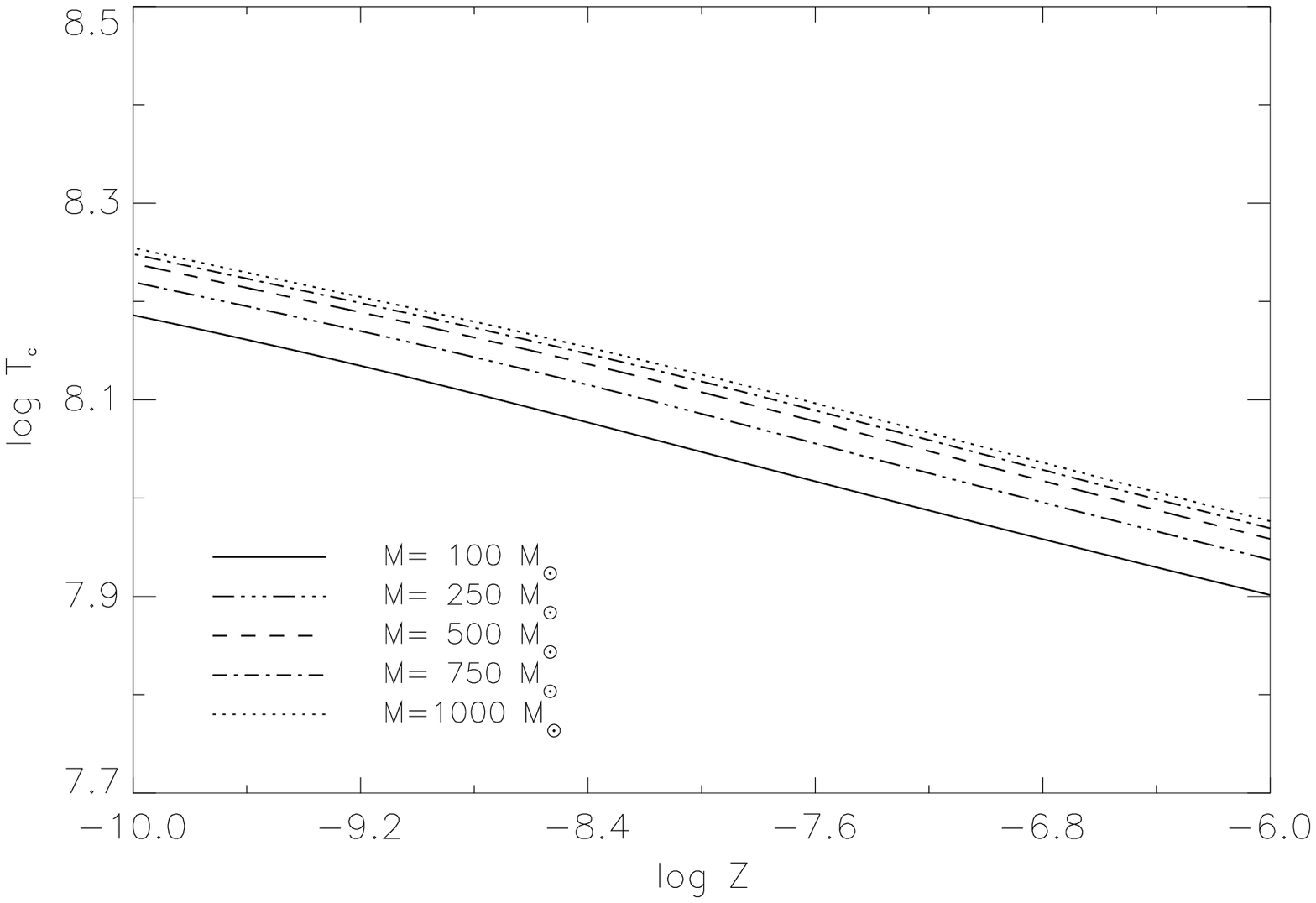}
%  \vspace*{174pt}
\caption{\emph{Ibidem}. Central temperature.} \label{bar_fig13}
\end{center}
\end{figure}

\begin{figure}
\begin{center}
\includegraphics [width=84mm]{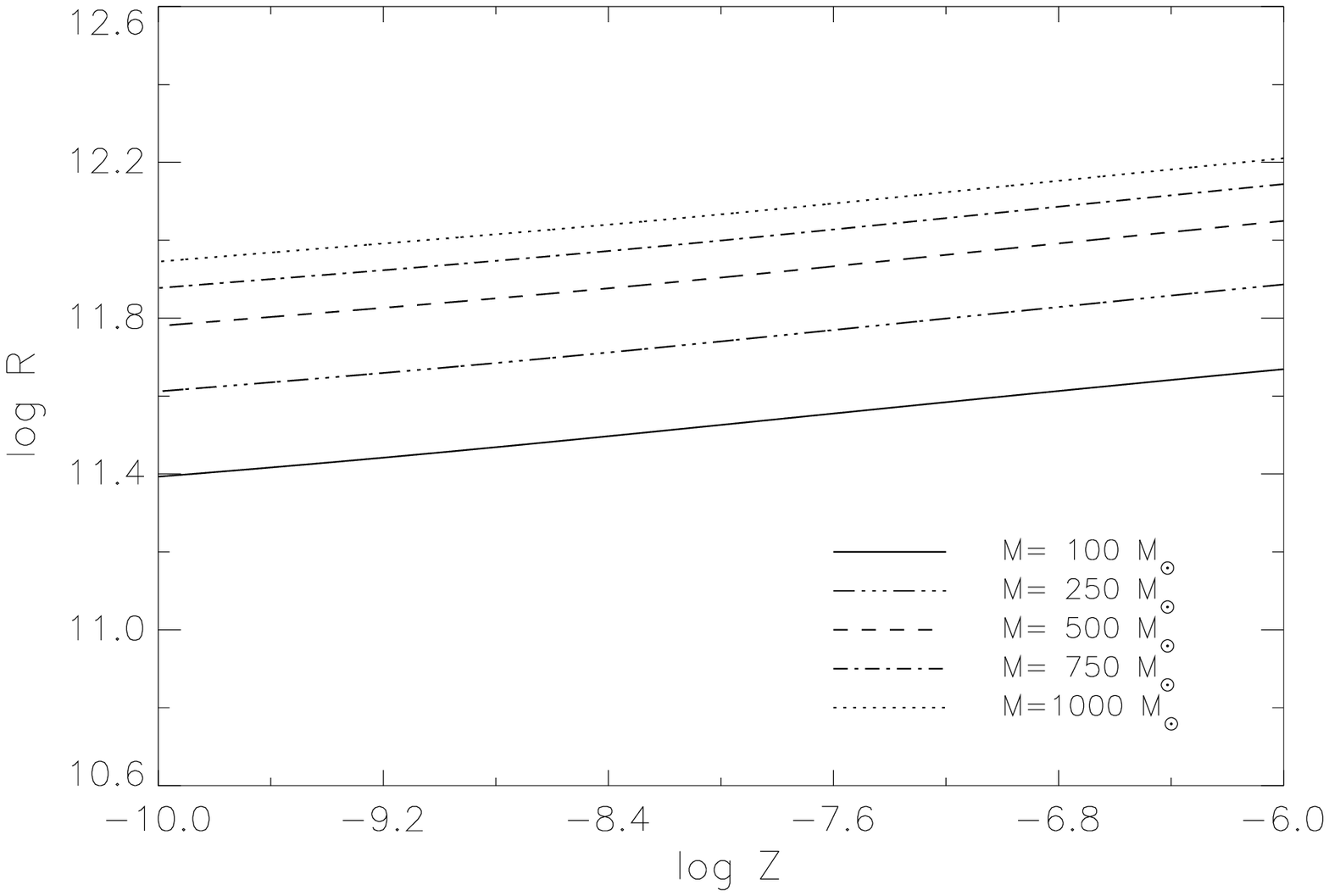}
%  \vspace*{174pt}
\caption{\emph{Ibidem}. Radius.} \label{bar_fig14}
\end{center}
\end{figure}

\begin{figure}
\begin{center}
\includegraphics [width=84mm]{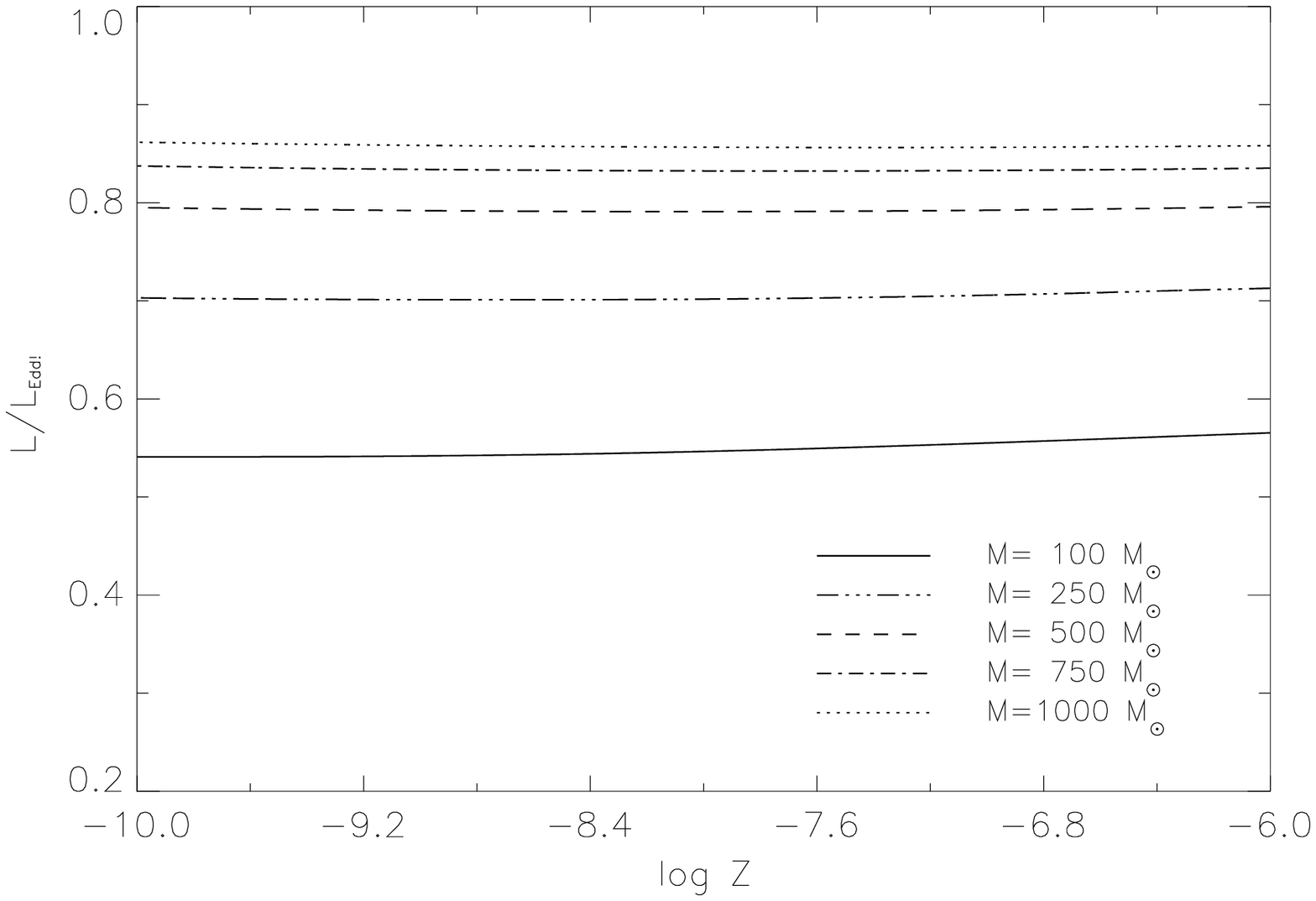}
%  \vspace*{174pt}
\caption{\emph{Ibidem}. Gamma factor.} \label{bar_fig15}
\end{center}
\end{figure}

\subsection{Stellar structure}

We have calculated the evolution of $100$, $150$, $200$, $250$ and
$300M_\odot$ stars with metallicities $Z=10^{-2}$, $10^{-3}$,
$10^{-6}$, $10^{-9}$, and $10^{-10}$. Pop III stars have higher
density, temperature and pressure than their Pop I and II
counterparts, and are radiation pressure dominated and very
luminous.

Pregalactic stars have central temperatures of about log $T_c\geq 8$
and effective temperatures of log $T_{eff} \leq 5$. A direct
consequence of higher central temperatures is that they have higher
energy generation rates. On the main sequence, lower metallicity
stars produce slightly less energy. However, these stars are hotter
than the others and so require higher temperatures to produce the
same amount of energy.

    The most striking feature of the low metallicity stellar
models is their atmospheric high temperature they are able to
maintain. During hydrogen burning, these stars derive their nuclear
energy from the inefficient pp-chains and the CNO-cycles. This is
possible because a small fraction of carbon is produced during the
pre-main sequence phase \citep{Castellani1983}. Some carbon is also
generated by the triple-$\alpha$ process before the star reaches the
main sequence \citep{Bromm2001}. That is, in the absence of metals
nuclear burning proceeds in a non-standard way. First, the hydrogen
burning occurs via the pp-chain. However, metal-free stars are
hotter and very luminous reaching high central temperatures which
are high enough for the simultaneous occurrence of helium burning
via the triple-$\alpha$ reaction. After a brief initial period of
this process, a trace amount of heavy metals are formed and this
makes that most of the energy generation rate during hydrogen
burning comes from the CNO-cycles.

    Lower metallicity stars have a higher central temperature to
support the star against gravity. This is a consequence of their
high central density and temperature, and because this type of stars
are very compact.

\subsection{Stellar evolution without mass loss}

\subsubsection{Physical variables}

For the present work stellar evolution models without mass loss have
been computed during the hydrogen and helium burning phases, for
very massive galactic and pregalactic Pop III $100, 150, 200, 250$
and $300 M_{\odot}$ stars with metallicities $Z=10^{-6}$ and
$10^{-9}$, respectively.

As an example of our evolution models, we list in Tables 1 and 2
some properties of $200 M_{\odot}$ galactic and pregalactic Pop III
stars.

The first two columns give the lifetime $\tau_{6}$, given in units
of $10^6$ years, and the helium mass fraction $X_{\rm He}$ during
the hydrogen and helium burning phases. Several physical variables
in logarithmic units are included such as the central density
$\rho_{\rm c}$, the central temperature $T_{\rm c}$, the luminosity
$L$ (in solar units), the effective temperature $T_{\rm eff}$, the
radius $R$, and the nuclear energy generation rate $\epsilon_{\rm
nuc}$. Then, the other quantities listed are the convective core
size $q_{\rm cc}$, which is the ratio of the mass of the convective
core to the total mass of the star, the radiation factor $\beta_{\rm
c}$ in the centre of the star, defined by $1-\beta =P_{r}/P$, where
$P_{r}$ is the radiation pressure and $P$ the total pressure, and
the Eddington luminosity factor $\Gamma$.

\begin{table*}
\small \caption{Physical variables and quantities for $200
M_{\odot}$ galactic Pop III stars with initial metallicity
$Z=10^{-6}$, without mass loss and with no-rotation, during the
hydrogen and helium burning phases.\label{tbl-1}}
\begin{tabular}{@{}ccccccccccc@{}}
\tableline $\tau_{6}$ & $X_{\rm He}$ & log $\rho_{\rm c}$ & log
$T_{\rm c}$ & log $\displaystyle{\frac{L}{L_{\odot}}}$ & log $T_{\rm
eff}$ & log $R$ & log $\epsilon_{\rm nuc}$ & $q_{\rm cc}$ &
$\beta_{\rm c}$ &
$\Gamma$ \\
\tableline

0.04407 & 0.23500 & 0.73760 & 7.87567 & 6.55521 & 4.87844 & 11.88977 & 5.76058 & 0.89348 & 0.42004 & 0.73286 \\
0.34457 & 0.30006 & 0.74533 & 7.88046 & 6.57138 & 4.87399 & 11.90675 & 5.77809 & 0.89744 & 0.40450 & 0.74736 \\
0.76005 & 0.40003 & 0.76022 & 7.88834 & 6.59460 & 4.86358 & 11.93919 & 5.80563 & 0.90758 & 0.38060 & 0.77037 \\
1.13125 & 0.50000 & 0.77986 & 7.89717 & 6.61619 & 4.84854 & 11.98007 & 5.83358 & 0.91887 & 0.35667 & 0.78979 \\
1.46357 & 0.60006 & 0.80543 & 7.90722 & 6.63618 & 4.82777 & 12.03159 & 5.86349 & 0.92563 & 0.33274 & 0.81376 \\
1.76063 & 0.70001 & 0.84086 & 7.91962 & 6.65458 & 4.80000 & 12.09633 & 5.89558 & 0.93542 & 0.30890 & 0.83259 \\
2.02594 & 0.80008 & 0.89235 & 7.93622 & 6.67161 & 4.76299 & 12.17888 & 5.93093 & 0.94607 & 0.28509 & 0.85120 \\
2.26415 & 0.90014 & 0.97789 & 7.96280 & 6.68784 & 4.71413 & 12.28471 & 5.96946 & 0.95284 & 0.26120 & 0.87436 \\
2.46034 & 0.99983 & 1.34155 & 8.07854 & 6.70443 & 4.67744 & 12.36639 & 5.97202 & 0.46767 & 0.23943 & 0.87764 \\
2.46037 & 0.99996 & 1.34222 & 8.07873 & 6.70444 & 4.67745 & 12.36637 & 5.97430 & 0.46771 & 0.23942 & 0.87766 \\
2.46045 & 0.99986 & 1.34278 & 8.07892 & 6.70445 & 4.67745 & 12.36638 & 5.97639 & 0.46776 & 0.23942 & 0.87784 \\
2.50534 & 0.90010 & 2.45172 & 8.42220 & 6.73086 & 4.43326 & 12.86797 & 6.41240 & 0.30804 & 0.26984 & 0.99198 \\
2.52450 & 0.80004 & 2.40005 & 8.40679 & 6.73326 & 4.32939 & 13.07691 & 6.39498 & 0.40993 & 0.26280 & 0.91068 \\
2.54783 & 0.70015 & 2.37961 & 8.40012 & 6.73469 & 4.24234 & 13.25172 & 6.39498 & 0.43042 & 0.25803 & 0.89312 \\
2.57361 & 0.60009 & 2.37056 & 8.39699 & 6.73547 & 4.17638 & 13.38405 & 6.39740 & 0.43887 & 0.25356 & 0.88887 \\
2.60105 & 0.50017 & 2.36860 & 8.39610 & 6.73588 & 4.14063 & 13.45575 & 6.40088 & 0.44543 & 0.24923 & 0.89512 \\
2.62969 & 0.40017 & 2.37269 & 8.39710 & 6.73603 & 4.13471 & 13.46766 & 6.40352 & 0.45009 & 0.24500 & 0.89164 \\
2.65913 & 0.30001 & 2.38339 & 8.40015 & 6.73596 & 4.14619 & 13.44467 & 6.40307 & 0.45347 & 0.24090 & 0.89055 \\
2.68881 & 0.20014 & 2.40316 & 8.40603 & 6.73581 & 4.15927 & 13.41843 & 6.40012 & 0.45335 & 0.23697 & 0.88681 \\
2.71795 & 0.10005 & 2.44120 & 8.41767 & 6.73569 & 4.16384 & 13.40923 & 6.36450 & 0.45693 & 0.23337 & 0.88916 \\
2.73912 & 0.01070 & 2.51746 & 8.44134 & 6.73582 & 4.15837 & 13.42023 & 5.91299 & 0.45401 & 0.23109 & 0.88956 \\
2.74053 & 0.00123 & 2.53190 & 8.44585 & 6.73588 & 4.15804 & 13.42092 & 5.50297 & 0.45446 & 0.23099 & 0.89017 \\
 \tableline
\end{tabular}
\end{table*}

\begin{table*}
\small \caption{Physical variables and quantities for $200
M_{\odot}$ pregalactic Pop III stars with initial metallicity
$Z=10^{-9}$, without mass loss and with no-rotation, during the
hydrogen and helium burning phases. \label{tbl-2}}
\begin{tabular}{@{}ccccccccccc@{}}
\tableline $\tau_{6}$ & $X_{\rm He}$ & log $\rho_{\rm c}$ & log
$T_{\rm c}$ & log $\displaystyle{\frac{L}{L_{\odot}}}$ & log $T_{\rm
eff}$ & log $R$ & log $\epsilon_{\rm nuc}$ & $q_{\rm cc}$ &
$\beta_{\rm c}$ &
$\Gamma$ \\
\tableline

0.04407 & 0.23500 & 1.46281 & 8.11724 & 6.56930 & 5.00039 & 11.65293 & 5.68780 & 0.87527 & 0.42049 & 0.73962 \\
0.32884 & 0.30008 & 1.47318 & 8.12283 & 6.58481 & 4.99569 & 11.67006 & 5.70573 & 0.88415 & 0.40503 & 0.75446 \\
0.72025 & 0.40000 & 1.49258 & 8.13204 & 6.60693 & 4.98487 & 11.70278 & 5.73226 & 0.75495 & 0.38138 & 0.77484 \\
1.06814 & 0.50003 & 1.51732 & 8.14230 & 6.62713 & 4.96938 & 11.74386 & 5.75997 & 0.90139 & 0.35781 & 0.79531 \\
1.37810 & 0.60005 & 1.54912 & 8.15406 & 6.64545 & 4.94843 & 11.79492 & 5.78962 & 0.64234 & 0.33438 & 0.81714 \\
1.65450 & 0.70003 & 1.59211 & 8.16853 & 6.66204 & 4.92085 & 11.85836 & 5.82162 & 0.91933 & 0.31113 & 0.83462 \\
1.90105 & 0.80005 & 1.65303 & 8.18772 & 6.67699 & 4.88498 & 11.93759 & 5.85683 & 0.93180 & 0.28799 & 0.85101 \\
2.11975 & 0.90016 & 1.75252 & 8.21821 & 6.69054 & 4.83938 & 12.03556 & 5.89503 & 0.93966 & 0.26493 & 0.87007 \\
2.30096 & 0.99992 & 2.19813 & 8.35947 & 6.70693 & 4.80493 & 12.11265 & 5.73874 & 0.41750 & 0.24530 & 0.85123 \\
2.30098 & 1.00006 & 2.19971 & 8.35997 & 6.70695 & 4.80505 & 12.11242 & 5.74387 & 0.41763 & 0.24533 & 0.85064 \\
2.30103 & 0.99994 & 2.20122 & 8.36047 & 6.70697 & 4.80517 & 12.11221 & 5.74887 & 0.41776 & 0.24534 & 0.85011 \\
2.32034 & 0.90054 & 2.40066 & 8.42261 & 6.71089 & 4.80396 & 12.11658 & 6.36513 & 0.41280 & 0.24675 & 0.88169 \\
2.33776 & 0.80019 & 2.35674 & 8.40641 & 6.71097 & 4.78656 & 12.15143 & 6.34291 & 0.42242 & 0.24439 & 0.87961 \\
2.35934 & 0.70013 & 2.34010 & 8.39941 & 6.71137 & 4.77235 & 12.18005 & 6.34426 & 0.42915 & 0.24187 & 0.87765 \\
2.38348 & 0.60006 & 2.33432 & 8.39608 & 6.71175 & 4.76042 & 12.20409 & 6.34729 & 0.43179 & 0.23922 & 0.87840 \\
2.40920 & 0.50011 & 2.33553 & 8.39510 & 6.71207 & 4.75027 & 12.22454 & 6.35226 & 0.43465 & 0.23647 & 0.88157 \\
2.43607 & 0.40027 & 2.34233 & 8.39598 & 6.71239 & 4.74112 & 12.24300 & 6.35622 & 0.43754 & 0.23367 & 0.88169 \\
2.46378 & 0.30015 & 2.35559 & 8.39895 & 6.71276 & 4.73206 & 12.26131 & 6.35760 & 0.44051 & 0.23084 & 0.88270 \\
2.49173 & 0.20016 & 2.37737 & 8.40467 & 6.71329 & 4.72292 & 12.27985 & 6.34892 & 0.43961 & 0.22803 & 0.88389 \\
2.51901 & 0.10025 & 2.41608 & 8.41584 & 6.71408 & 4.71410 & 12.29789 & 6.31307 & 0.43980 & 0.22538 & 0.88633 \\
2.53894 & 0.01019 & 2.48863 & 8.43781 & 6.71513 & 4.70969 & 12.30723 & 5.81421 & 0.43147 & 0.22366 & 0.88707 \\
2.53997 & 0.00157 & 2.49814 & 8.44068 & 6.71524 & 4.70981 & 12.30705 & 5.86620 & 0.43179 & 0.22360 & 0.88798 \\

\tableline
\end{tabular}
\end{table*}

These tables summarize the most representative data of the models,
and they include hydrogen and helium burning from the initial
chemical composition, in $10$\% intervals of the helium mass
fraction, to the end of helium burning when the helium mass fraction
is approximately $0.01$.

VMS are hot and luminous and so are located in the left upper part
of the Hertzprung-Russell (HR) diagram. Because Pop III stars are
hotter than their enriched counterparts, their locus in the
HR-diagram is shifted to the left upper part; pregalactic stars are
bluer than galactic ones.

For a given stellar mass, the evolution of a massive star, i.e., its
location in the HR-diagram, depends strongly on metallicity.
Metal-free stars have unique physical characteristics and they
exhibit high effective temperatures and small radii. In relationship
with their cosmological consequences, metal-free models are
important for predicting the ionizing photon production of the first
generation of stars.

Pregalactic stars were denser and hotter than galactic Pop III
stars. Some of their main physical variables during the hydrogen and
helium burning phases are shown in Figs.\ref{bar_fig15} to 21.

\begin{figure}
\begin{center}
\includegraphics [width=84mm]{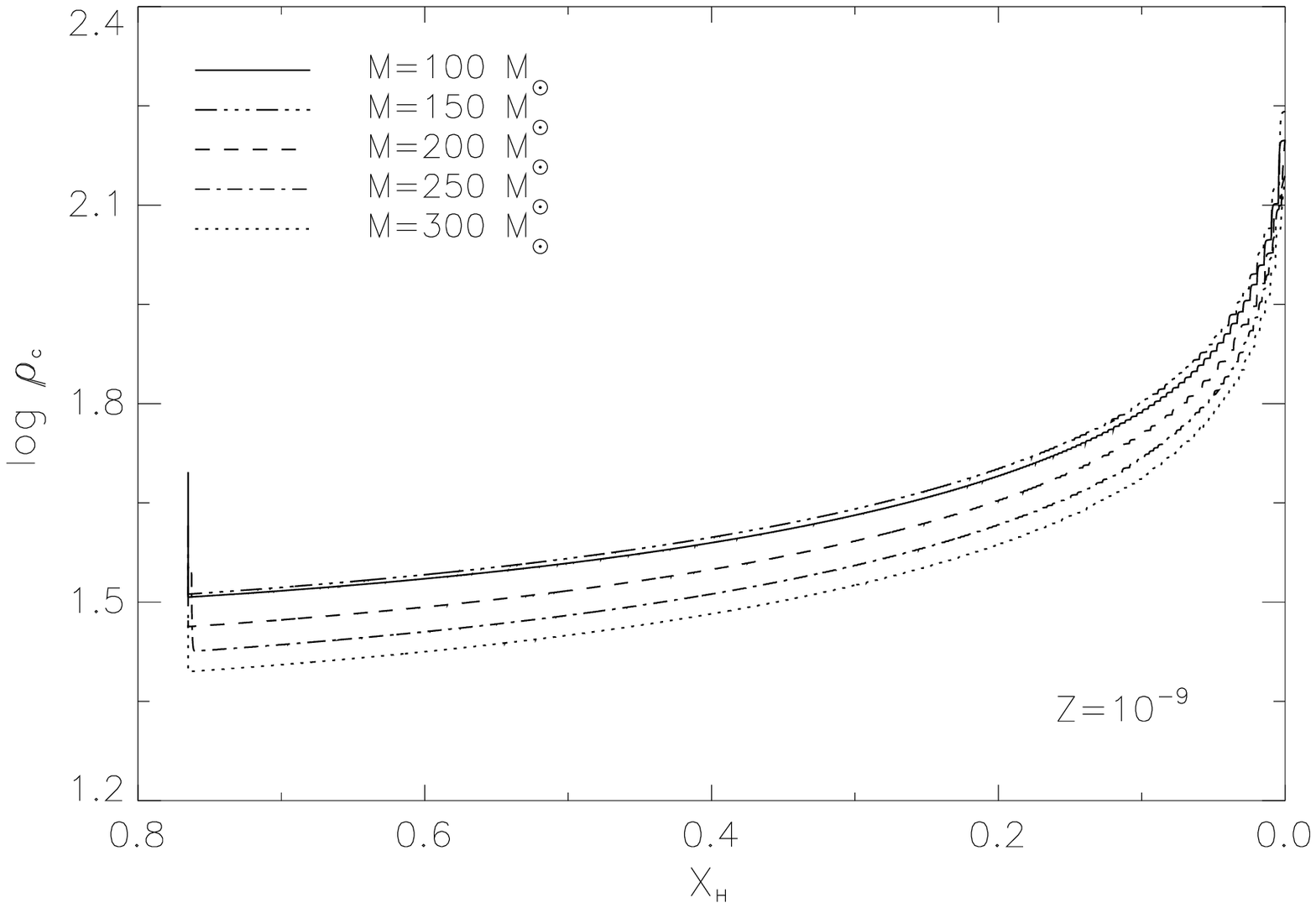}
%  \vspace*{174pt}
\caption{Central density for $100 M_\odot$ (solid line), $150
M_\odot$ (dash-dot-dot-dot-dash), $200 M_\odot$ (dashes), $250
M_\odot$ (dash-dot-dash) and $300 M_\odot$ (dots) pregalactic Pop
III stars with metallicity $Z=10^{-9}$, without mass loss, during
the hydrogen burning.} \label{bar_fig16}
\end{center}
\end{figure}

\begin{figure}
\begin{center}
\includegraphics [width=84mm]{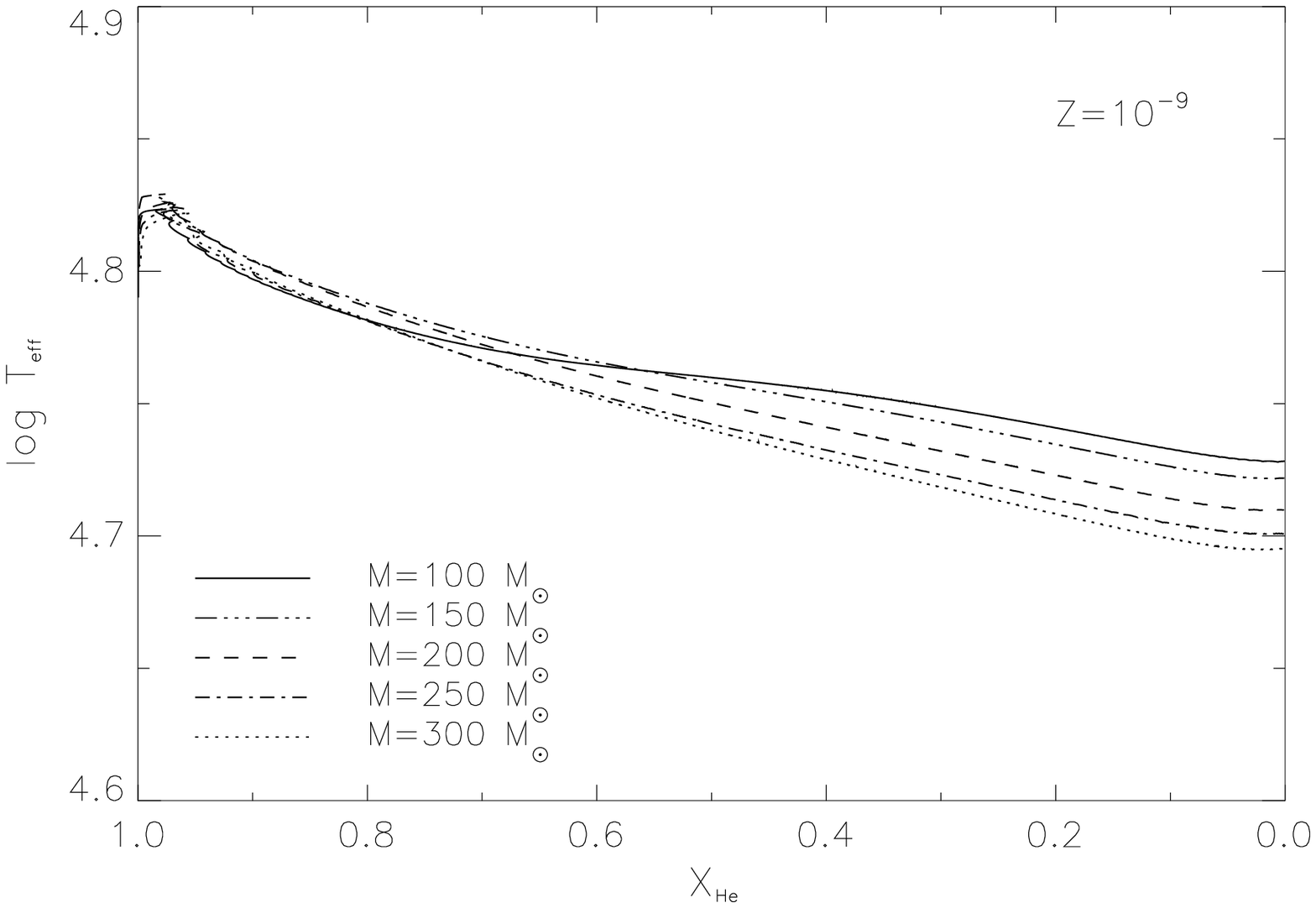}
%  \vspace*{174pt}
\caption{\emph{Ibidem}. Effective temperature.} \label{bar_fig17}
\end{center}
\end{figure}

\begin{figure}
\begin{center}
\includegraphics [width=84mm]{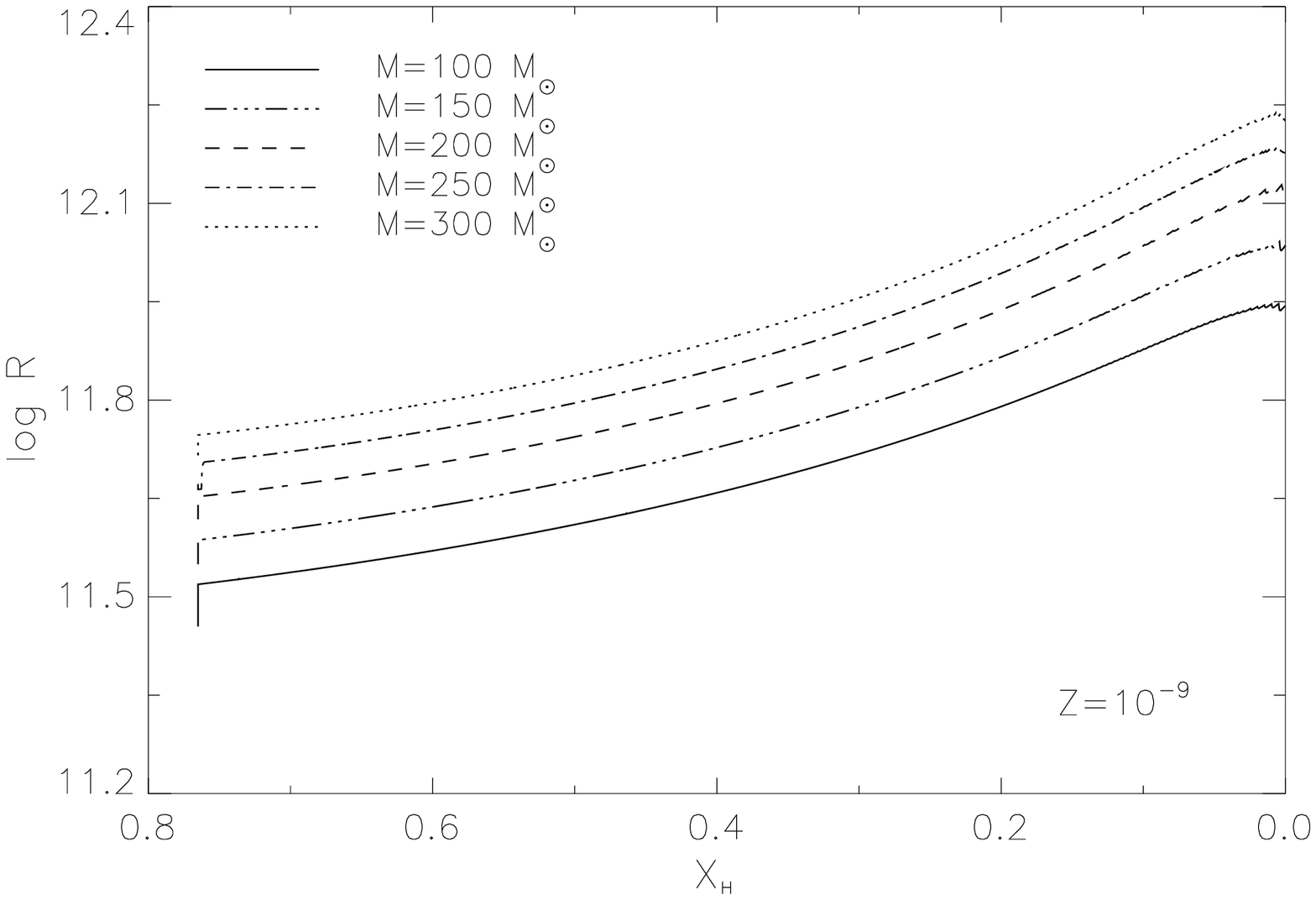}
%  \vspace*{174pt}
\caption{\emph{Ibidem}. Radius.} \label{bar_fig18}
\end{center}
\end{figure}

\begin{figure}
\begin{center}
\includegraphics [width=84mm]{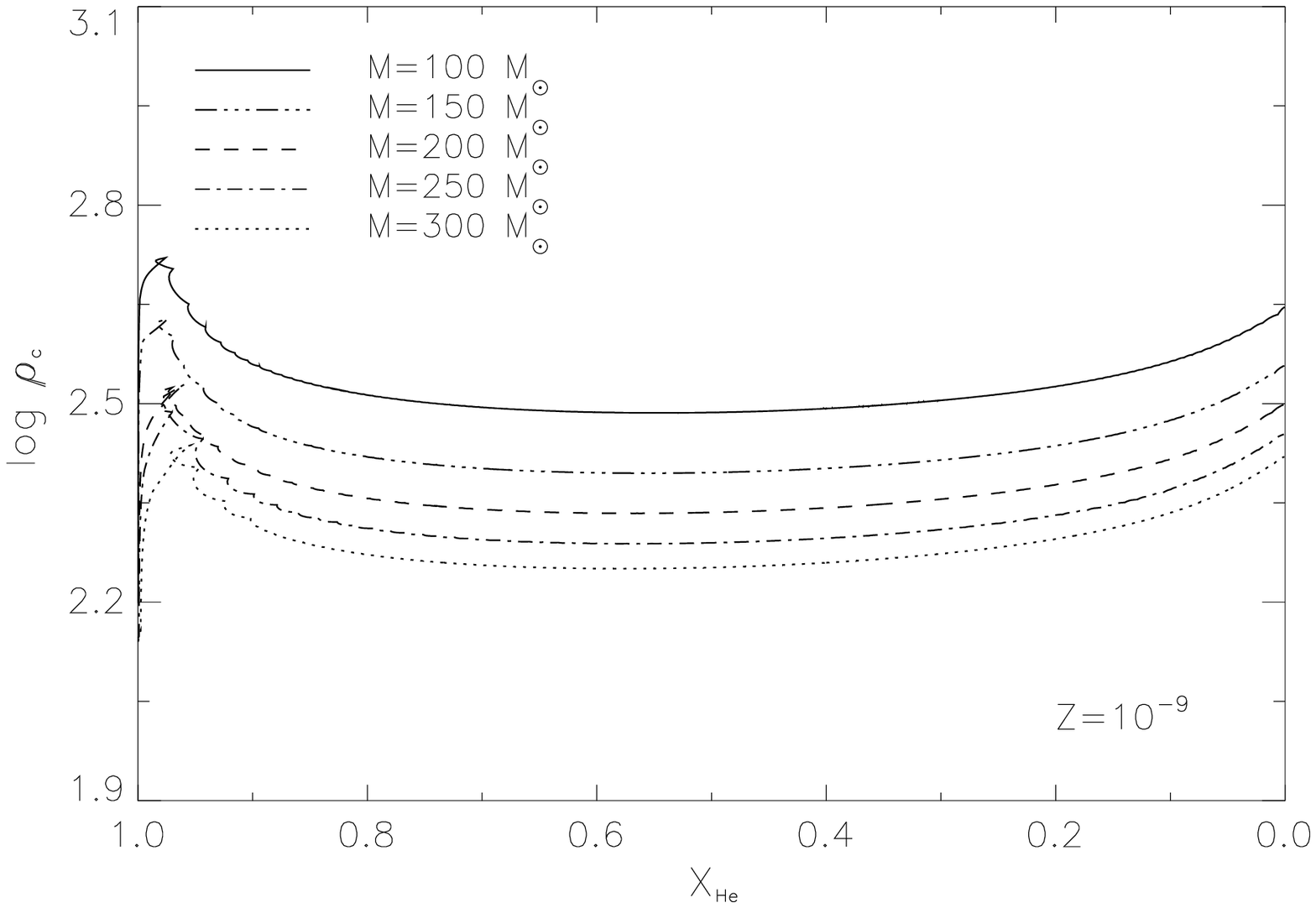}
%  \vspace*{174pt}
\caption{Central density for $100 M_\odot$ (solid line), $150
M_\odot$ (dash-dot-dot-dot-dash), $200 M_\odot$ (dashes), $250
M_\odot$ (dash-dot-dash) and $300 M_\odot$ (dots) pregalactic Pop
III stars with metallicity $Z=10^{-9}$, without mass loss, during
the helium burning phase.} \label{bar_fig19}
\end{center}
\end{figure}

\begin{figure}
\begin{center}
\includegraphics [width=84mm]{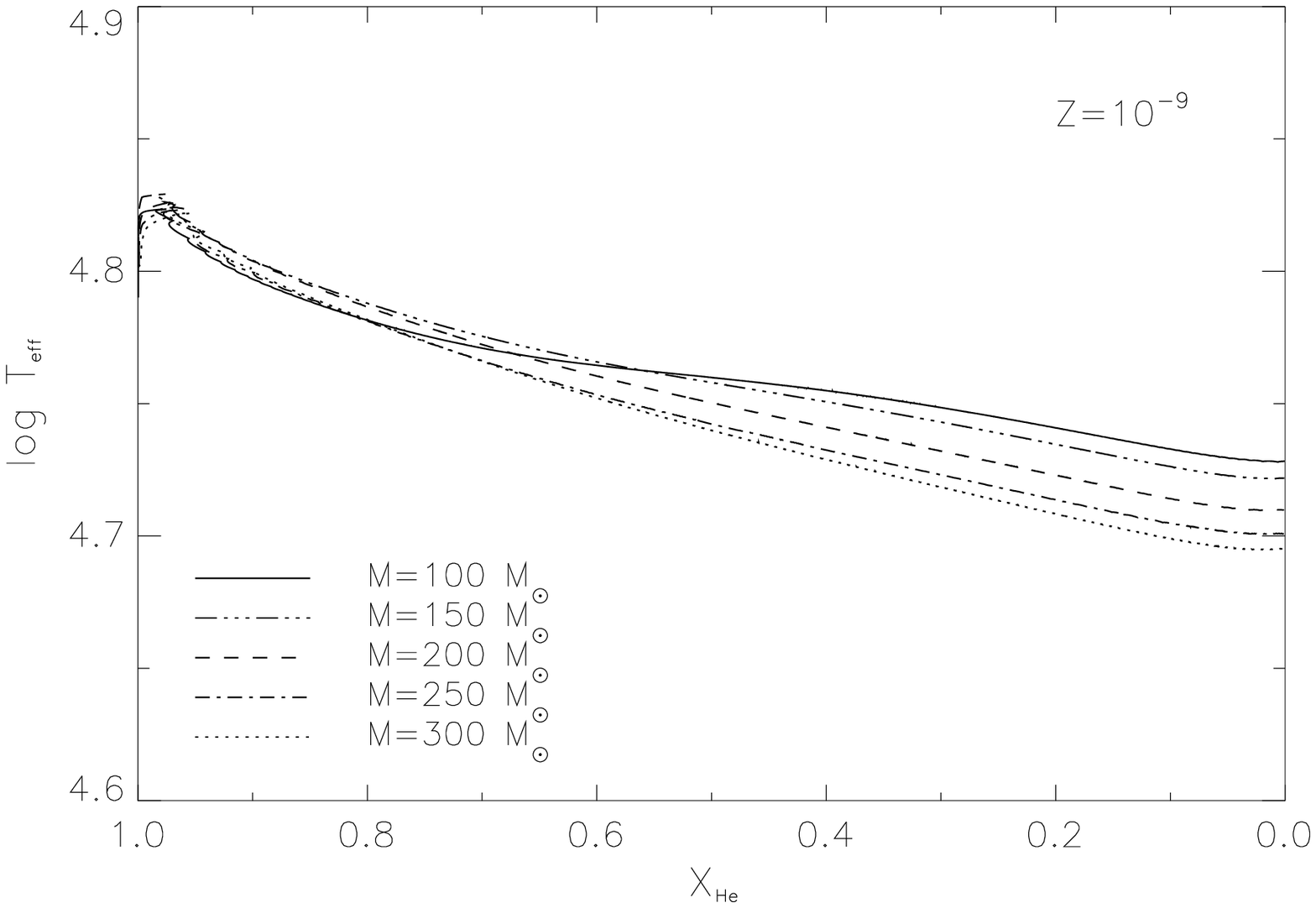}
%  \vspace*{174pt}
\caption{\emph{Ibidem}. Effective temperature.} \label{bar_fig20}
\end{center}
\end{figure}

\begin{figure}
\begin{center}
\includegraphics [width=84mm]{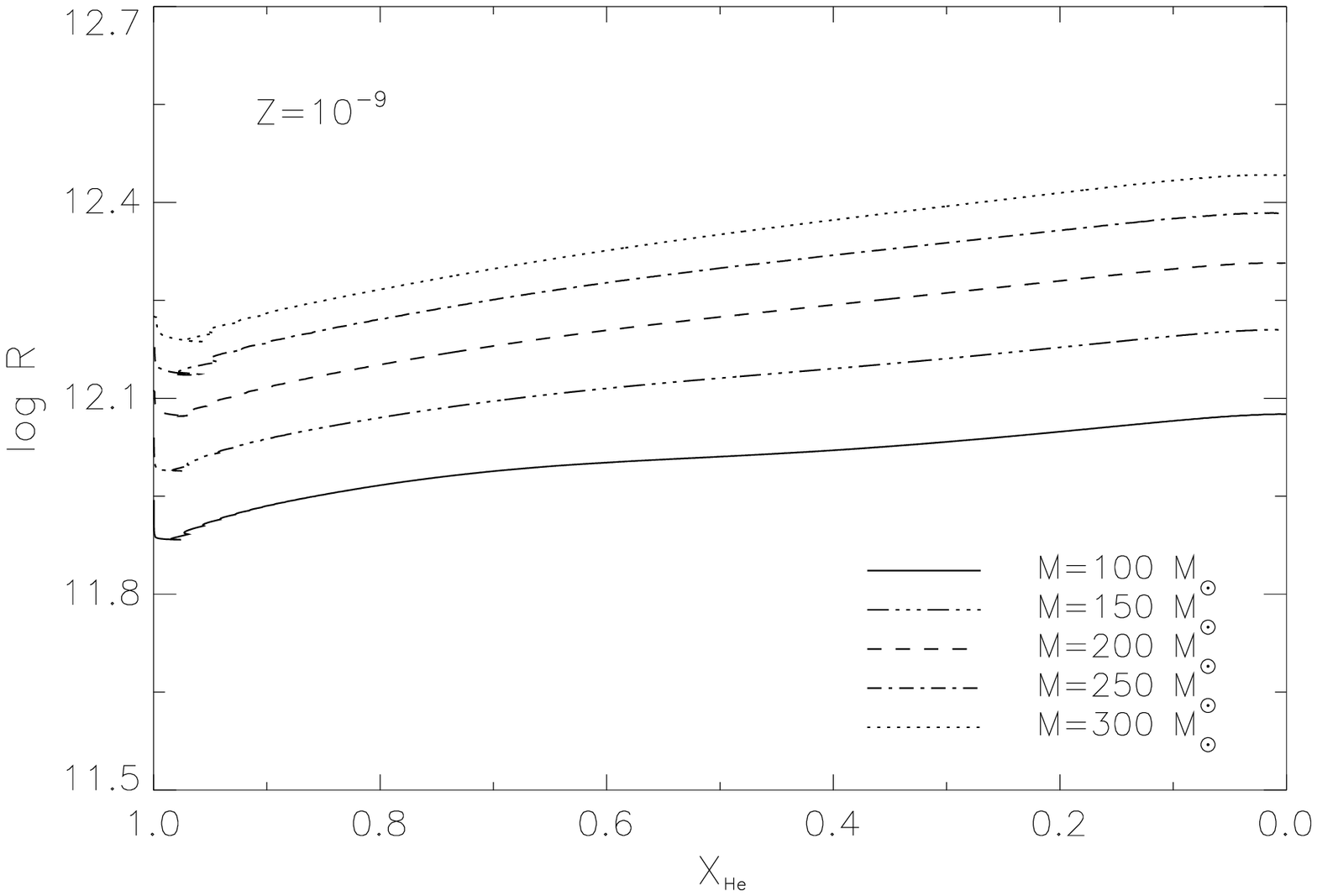}
%  \vspace*{174pt}
\caption{\emph{Ibidem}. Radius.} \label{bar_fig21}
\end{center}
\end{figure}

The following are the main properties of the studied evolutionary
models for galactic and pregalactic very massive Pop III stars.

\subsubsection*{\qquad a) Central density}

Pop III stars with $Z=10^{-6}$ settle down on the main sequence with
central densities of $\rho_{\rm c}=7.46$, $5.97$ and $3.94$ g
cm$^{-3}$ for $100$, $200$ and $300 M_{\odot}$, respectively. For
the same stellar masses, and $Z=10^{-9}$, central densities are
$\rho_{\rm c}=32.18$, $29.03$ and $24.83$ g cm$^{-3}$, respectively.
That is, the central density is higher for lower metallicity stars,
and so, Pop III pregalactic stars are denser than galactic stars.
However, the central density decreases with increasing stellar mass.

During hydrogen burning, the central density increases slowly but
then by the end of this burning phase its increase rate grow
significantly. The central density increases during the transition
from hydrogen to helium burning and keeps increasing during the
whole helium burning phase.

\subsubsection*{\qquad b) Central temperature}

The VMS high energy requirements, demand high central temperatures
in order to be able to maintain their structure and energy output.
That is, the central temperature is high for more massive stars, but
with lower metallicity the stars have an even higher temperature. At
the beginning of hydrogen burning, the central temperature with
$Z=10^{-6}$ is $T_{7}=7.02$, $7.51$ and $7.76$, for $100$, $200$ and
$300 M_{\odot}$ stars, respectively, where $T_{n}=T/10^{n}$ K. For
$Z=10^{-9}$, the corresponding temperatures are $T_{8}=1.14$, $1.30$
and $1.36$, respectively. During hydrogen burning, $T_{\rm c}$
slowly increases until the end of this burning phase when it begins
to increase to high values and continues to do so during the whole
helium burning phase.

\subsubsection*{\qquad c) Nuclear energy generation}

During hydrogen burning the main energy source is given by the
CNO-cycles because of their strong temperature dependence. In lower
metallicity stars these cycles are activated just after a short
helium burning phase that produces enough CNO elements for the
CNO-cycles to operate.

With increasing stellar mass, the central temperature is higher and
so the nuclear energy generation. However, for pregalactic stars it
increases with decreasing metallicity. As a consequence of the
larger central temperatures, CNO-cycles carbon production is
enhanced at an earlier evolutionary phase than in less massive
stars. The nuclear energy generation increases during hydrogen
burning. But, for galactic Pop III stars with metallicity
$Z=10^{-6}$, the energy generation rate decreases when hydrogen
tends to be exhausted. When the existing hydrogen mass fraction is
about $\sim 0.02$, and the helium mass fraction reaches $\sim 0.98$,
the nuclear energy generation decreases. For pregalactic stars with
$Z=10^{-9}$, during the transition from hydrogen to helium burning
there is only a small decrease in the nuclear energy generation
rate, i.e. the transition from hydrogen to helium burning is smooth,
quite different than for galactic stars.

There are different nuclear generation rates at the end of hydrogen
burning because the central temperature in galactic Pop III stars is
not high enough for helium ignition. This situation does not take
place in the case of pregalactic stars precisely because they have a
higher central temperature at the end of hydrogen burning. Then, the
transition to the next burning phase occurs in a smoother form. In
the other case, a strong explosive helium flash takes place during
the transition to helium burning. At this moment, the star is
contracting, consequently, it is heated and reaches an appropriate
central temperature to ignite helium.

When helium is ignited in the core of the star, the nuclear energy
generation rate increases rapidly during the transition to helium
burning, then reaches a maximum and decreases towards the end of
this burning phase. This occurs for galactic stars with metallicity
$Z=10^{-6}$. In the case of pregalactic stars, with metallicity
$Z=10^{-9}$, this transition takes place very smoothly because the
stars are hot enough to ignite helium immediately after exhausting
hydrogen. Then the generation rate increases towards a maximum and
decreases by the end of helium burning.

\subsubsection*{\qquad d) Luminosity}

Very massive Pop III stars are very luminous. For $100$, $200$ and
$300 M_{\odot}$ with metallicity $Z=10^{-6}$, at the beginning of
the main sequence, their luminosity are $L=1.22\times 10^{6}$,
3.59$\times 10^{6}$ and $6.31\times 10^{6} L_{\odot}$, respectively.
For stars with $Z=10^{-9}$, the corresponding figures are
$L=1.27\times 10^{6}$, $3.71\times 10^{6}$ and $6.51\times 10^{6}
L_{\odot}$. During hydrogen burning the stellar luminosity increases
slightly and varies smoothly during the transition from hydrogen to
helium burning. Then it remains practically constant during helium
burning.

\subsubsection*{\qquad e) Effective temperature}

Pregalactic Pop III stars are hotter than galactic stars. At the
beginning of the main sequence, galactic stars with $Z=10^{-6}$ have
an effective temperature $T_{\rm eff}=69582$, $75586$ and $77988$ K
for $100$, $200$ and $300 M_{\odot}$, respectively. For the same
stellar masses, pregalactic stars with $Z=10^{-9}$ have $T_{\rm
eff}=89170$, $100090$ and $103390$ K, respectively. These different
values are due to their higher central temperatures and the
different mechanisms to drive nuclear burning.

The effective temperature continuously decrease during hydrogen
burning, until the transition to helium burning that it increases.
Then, the effective temperature start to decrease but then remains
practically constant until the end of this burning phase when the
effective temperature decreases even more.

\subsubsection*{\qquad f) Radius}

Very massive Pop III stars are compact. On the main sequence,
galactic stars with metallicity $Z=10^{-6}$ have initial radii
$R=7.67$, $11.15$ and $13.88 R_{\odot}$ for $100$, $200$ and $300
M_{\odot}$, respectively. For pregalactic stars with $Z=10^{-9}$,
their radii are $R=4.75$, $6.46$ and $8.02 R_{\odot}$, respectively.

As the effective temperature decreases, the radii increases. At the
end of hydrogen burning, for galactic $100$, $200$ and $300
M_{\odot}$ stars with initial metallicity $Z=10^{-6}$ their radii
are $R=20.82$, $33.40$ and $43.39 R_{\odot}$, respectively; and for
pregalactic stars with initial metallicity $Z=10^{-9}$, $R=12.62$,
$18.61$ and $24.18 R_{\odot}$, respectively.

On the contrary to effective temperatures, during the transition
from hydrogen to helium burning, radii slowly increase and remain
almost constant during helium burning until the end when they
increase again. This is, pregalactic lower metallicity stars are
more compact because they are hotter than their galactic
counterparts.

\subsubsection*{\qquad g) Convective core}

A convective core is always present from the ZAMS to the end of the
helium burning phase. Very massive Pop III stars develop a large
convective core. For $100$, $200$ and $300 M_{\odot}$ stars with
$Z=10^{-6}$, the convective core size at the beginning of the main
sequence is $q_{\rm cc}=0.81$, $0.89$ and $0.92$, respectively. For
the same stellar masses and $Z=10^{-9}$, $q_{\rm cc}=0.79$, $0.88$
and $0.91$, respectively. The core is larger for higher metallicity.

The convective core size is larger for higher masses and increases
during hydrogen burning. In fact, the studied stars are almost fully
convective during hydrogen burning. At the end of this burning phase
the stars contract while forming a helium core. For the masses above
mentioned and metallicity $Z=10^{-6}$, at the end of hydrogen
burning, the convective core size $q_{\rm cc}=0.42$, $0.47$ and
$0.48$ for $100$, $200$ and $300 M_{\odot}$, respectively. For
$Z=10^{-9}$, $q_{\rm cc}=0.40$, $0.42$ and $0.44$, respectively.
Then, stars with $Z=10^{-6}$ form a helium core of $M_{\rm
He}=42.5$, $98.3$ and $142.7 M_{\odot}$; with $Z=10^{-9}$, the core
masses are $M_{\rm He}=39.7$, $93.5$ and $142.4 M_{\odot}$,
respectively.

Because for galactic stars with metallicity $Z=10^{-6}$ the
transition from hydrogen to helium burning is explosive, they
suddenly contract, affecting momentarily their core size which
rapidly increases, but then decreases while forming a helium core.
That is, the exhaustion of hydrogen in the centre of the star causes
a progressive contraction of the star and the shrinking of the
convective core which finally vanishes when $X_{c}\sim 10^{-3}$.
During the transition, the energy released by the star is supplied
by the gravitational contraction. In the pregalactic case with
$Z=10^{-9}$ the burning transition is very smooth and the stars
contract immediately forming a helium core.

During helium burning, for $100$, $200$ and $300 M_{\odot}$ stars
with $Z=10^{-6}$, the convective core is $q_{\rm cc}\sim 0.40$,
$0.45$ and $0.47$, respectively. For $Z=10^{-9}$, $q_{\rm cc}\sim
0.39$, $0.43$ and $0.45$, respectively. That is, a carbon core mass
of $M_{\rm C}=40.5$, $90.9$ and $130.6 M_{\odot}$, respectively, is
formed for stars with $Z=10^{-6}$, while for $Z=10^{-9}$, $M_{\rm
C}=39.1$, $86.4$ and $134.7 M_{\odot}$, respectively.

\subsubsection*{\qquad h) Radiation pressure}

Very massive Pop III stars are dominated by radiation pressure. At
the centre of the stars we have $1-\beta_{\rm c}=0.45$, $0.58$ and
$0.64$ for $100$, $200$ and $300 M_{\odot}$ galactic stars with
metallicity $Z=10^{-6}$. That is, the contribution of radiation
pressure to the total pressure is considerable. For pregalactic
stars with metallicity $Z=10^{-9}$ we have that $1-\beta_{\rm
c}=0.45$, $0.58$ and $0.64$, respectively.

The contribution of the central radiation pressure increases during
hydrogen burning. After the transition from hydrogen to helium
burning, the central radiation pressure contribution increases and
then remains almost constant.

\subsubsection*{\qquad i) The Eddington luminosity}

Very massive Pop III stars evolve during hydrogen burning below the
Eddington upper luminosity limit. For $Z=10^{-6}$, at the beginning
of hydrogen burning, the ratio $\Gamma=0.60$, $0.73$ and $0.79$ for
$100$, $200$ and $300 M_{\odot}$, respectively; and for $Z=10^{-9}$
$\Gamma=0.60$, $0.74$ and $0.80$, respectively.

During hydrogen burning, this ratio increases and has a maximum
close to the end of this burning phase, approximately when the
helium mass fraction is $0.98$ and then decreases slightly as the
star reaches helium ignition. After the transition from hydrogen to
helium burning, $\Gamma$ increases when the helium mass fraction is
approximately $0.98$. Furthermore, this ratio decreases slightly but
then remains almost constant during the helium burning phase.

\begin{figure*}
\begin{center}
\includegraphics [width=170mm] {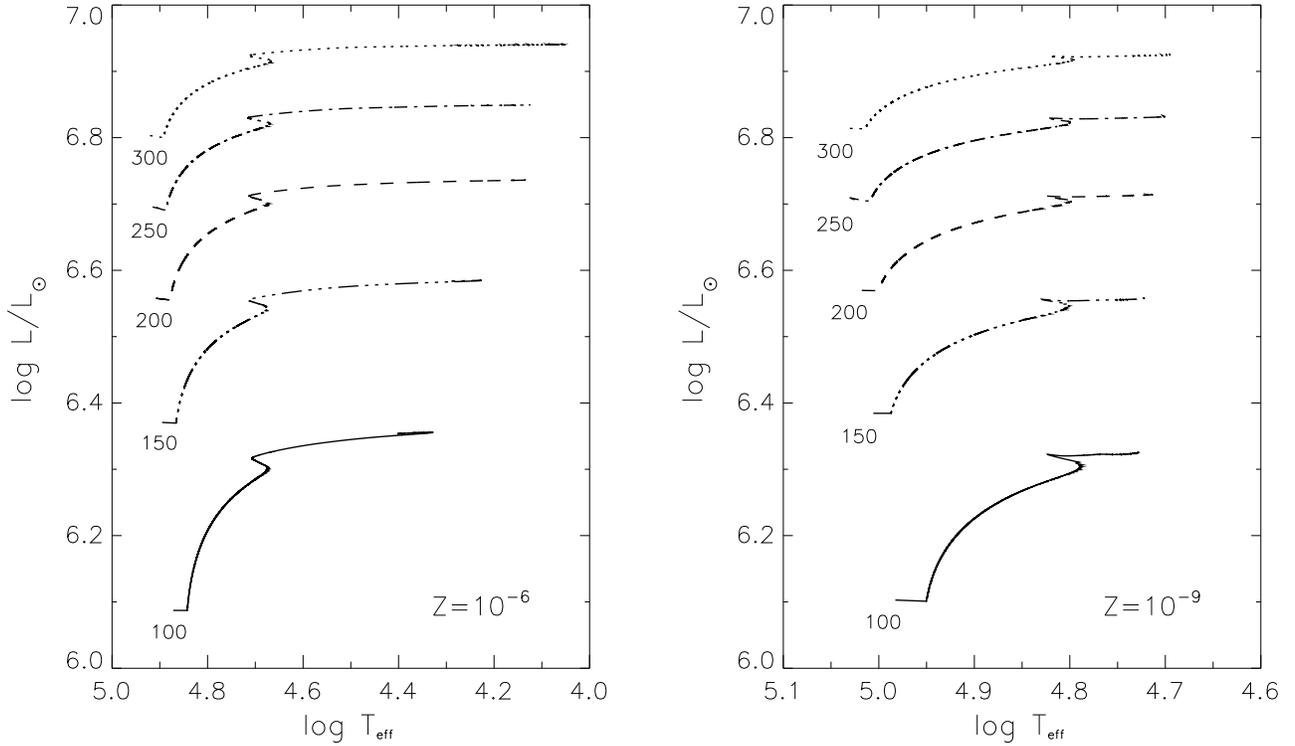}
\caption{Evolutionary tracks in the HR-diagram for $100$, $150$,
$200$, $250$ and $300 M_{\odot}$ Pop III stars with metallicity
$Z=10^{-6}$ and $Z=10^{-9}$, respectively and without mass loss
during the hydrogen and helium burning phases.} \label{bar_fig22}
\end{center}
\end{figure*}

\subsubsection{Evolutionary tracks}

Fig. \ref{bar_fig22} shows evolutionary tracks in the HR-diagram for
$100$, $150$, $200$, $250$ and $300 M_{\odot}$ galactic and
pregalactic Pop III stars with metallicity $Z=10^{-6}$ and
$10^{-9}$, respectively. The most massive stars are the hotter and
most luminous. Luminosity decreases with lower metallicity. For the
same stellar mass, galactic and pregalactic Pop III stars have
similar luminosities. All stars settle down on the main-sequence
with a high effective temperature and luminosity. During hydrogen
burning all stars increase their luminosity while the helium mass
fraction increase and both the central density and temperature
increase.

For stars with metallicity $Z=10^{-6}$, the transition between
nuclear burnings is explosive while the star contracts. At this
moment, the effective temperature and luminosity increase. Then, the
luminosity remains almost constant while the effective temperature
decreases. However, for different stars their luminosity and
effective temperature decrease with decreasing mass and metallicity.
For pregalactic stars the hydrogen to helium burning transition
occurs very smoothly because they are hotter and their central
temperature is high enough to ignite helium promptly.

Galactic and pregalactic stars evolve with different lifetimes. For
$100$, $200$ and $300 M_{\odot}$ galactic Pop III stars, their
lifetimes during hydrogen burning are $3.07566$, $2.42932$ and
$2.20822$ megayears, respectively. For pregalactic stars, their
lifetimes for the same masses are $2.87012$, $2.28397$ and $2.09287$
megayears, respectively. During helium burning, stellar lifetimes
are shorter than during hydrogen burning.

The initial hydrogen and helium burning phases take place at the
blue side of the HR-diagram. When the stars evolve they move toward
the red. However, the presently studied galactic and pregalactic
very massive Pop III stars, evolving without mass loss and
no-rotation, do not experience the asymptotic giant branch (AGB)
phase. The most massive zero-metal stars tend to be cooler but the
temperature remains above $\sim 10^{4}$ K.

The HR-diagram evolution of massive lower metallicity Pop III stars
is different than for Pop I and II stars, because they do not evolve
to red-giants before core collapse \citep{Umeda2000}. In the quoted
range of masses used by \citep{Castellani1983}, massive zero-metal
stars fail to reach the red giant region either at the hydrogen or
helium burning phases.

In the Hertzprung-Russell diagram the locus of very massive Pop III
stars is in the left upper part. These stars are hotter and very
luminous. Pregalactic stars are hotter than galactic Pop III stars.
Then, stars with lower metallicity are shifted to the left because
they are bluer than the others.

\subsubsection*{\qquad k) $\rho-T$ plane}

The $\rho-T$ plane describes the state of the gas in the innermost
stellar regions and it is important for the diagnosis of the stellar
structure and evolution. The evolution of the central conditions
determine the boundaries in which the equation of state is dominated
by different pressure components, i.e., ideal gas, radiation
pressure or electron pressure.

According to the zones of the equation of state of an electron gas,
the studied stars occupy the upper loci of a non-degenerate and
non-relativistic gas. In this zone there is a boundary at which pair
production could become important. Regarding the gas in
thermodynamic equilibrium, these stars are dominated by radiation
pressure.

The central conditions varies in the range log $\rho_{\rm c}\sim 1 -
3$ and log $T_{\rm c}=8.0-8.6$ for galactic stars with $Z=10^{-6}$.
For pregalactic stars with $Z=10^{-9}$, both the central temperature
and density are higher. The evolution of the central conditions can
be described by $T_{\rm c}\sim\rho_{\rm c}^{1/3}$. This behaviour
depends on the equation of state and it is similar for models of any
metallicity.

We distinguish two regions in the $\rho-T$ plane in correspondence
to the stages of gravitational contraction of stellar cores between
nuclear burnings. In the second region, corresponding to helium
burning, the central density and temperature increases more than
during hydrogen burning.

\section{Discussion}
\label{sec:discussion}

The evolutionary tracks of our massive Pop III stars are shifted to
the left upper part of the HR-diagram as in Tumlinson et al. (2003)
models. Then, stars evolve to the red with increasing luminosity and
decreasing effective temperature. In both cases, luminosities are
similar but, in the present case, final effective temperatures are
slightly higher. This is probably due to the different parameters
used, chemical composition and specific implementation of physical
processes, e.g., convection.

\subsubsection*{\qquad a) Nuclear lifetimes}

Our models are hot and luminous and have short lifetimes. For $100
M_{\odot}$ stars with $Z=10^{-6}$ and $10^{-9}$, their lifetimes
during hydrogen burning are $\tau_{\rm H}=3.09368$ and $2.92147$
megayears, respectively. Lifetimes during helium burning are of the
order of $\sim 10\%$ of their lifetime during hydrogen burning. Our
results for galactic and pregalactic hydrogen burning Pop III stars
are in good agreement with other authors.

For $M\gtrsim 300 M_{\odot}$, even metal-free stars evolve toward
$T_{\rm eff}<10^{4}$ K and eventually become red supergiants, as the
hydrogen burning shell becomes more active with increasing stellar
mass \citep{Baraffe2001}. In the present work, the most massive
metal-free stars tend to be cooler and likely could become red
supergiants. However, for a helium mass fraction equal to $\sim
0.01$, they maintain high effective temperatures. In all cases,
stars do not reach low effective temperatures. Evolving stars
without mass loss and with no-rotation, fail to reach the AGB phase.
At the end of helium burning, the studied galactic stars are hotter
than log $T_{\rm eff}\sim$ $4.0$ and pregalactic stars are hotter
than $T_{\rm eff}\sim$ $4.6$.

\subsubsection*{\qquad b) Nuclear energy generation}

The peculiar behaviour of low metallicity stars was first pointed
out by Ezer (1961) and their structure during hydrogen and helium
burning has been investigated by several authors.

According to the results presented here the stars begin to settle
down on the main-sequence with higher initial central temperatures
of order log $T_{\rm c}\sim$ $8$. Then, the onset of the $3\alpha$
reaction occurs at earlier stages of hydrogen burning. This is
because the $3\alpha$ reaction requires a much higher temperature
than the \emph{pp}-chains and ignites at the beginning of the
hydrogen burning phase. In the present models, after a brief initial
period of $3\alpha$ burning, a trace amount of heavy elements has
been formed. Then, the stars expand and follow the CNO-cycles.

\begin{figure*}
\begin{center}
\includegraphics [width=164mm]{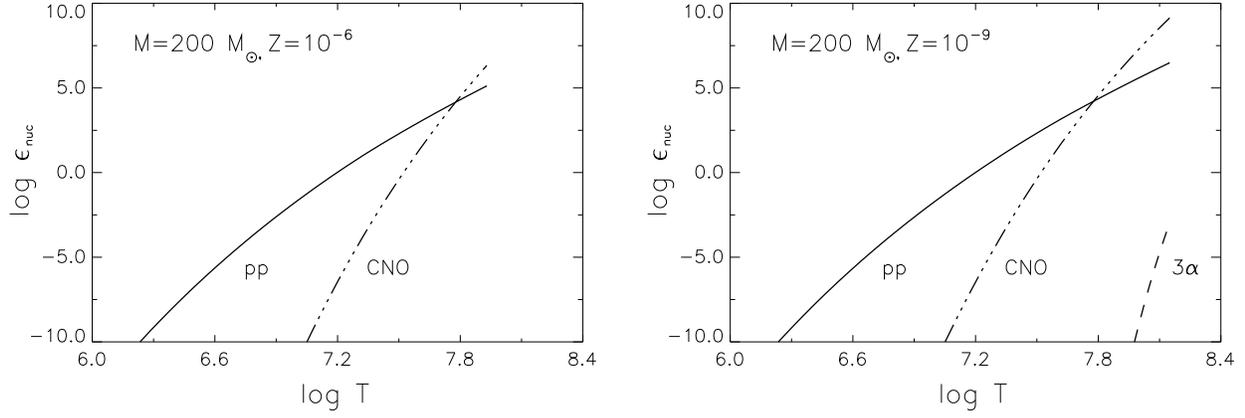}
%  \vspace*{174pt}
\caption{Main sequence nuclear energy generation rate as function of
the temperature for $200 M_{\odot}$ galactic (left panel), and
pregalactic (right panel) Pop III stars, with metallicity
$Z=10^{-6}$ and $Z=10^{-9}$, respectively.} \label{bar_fig23}
\end{center}
\end{figure*}

\begin{figure*}
\begin{center}
\includegraphics [width=164mm]{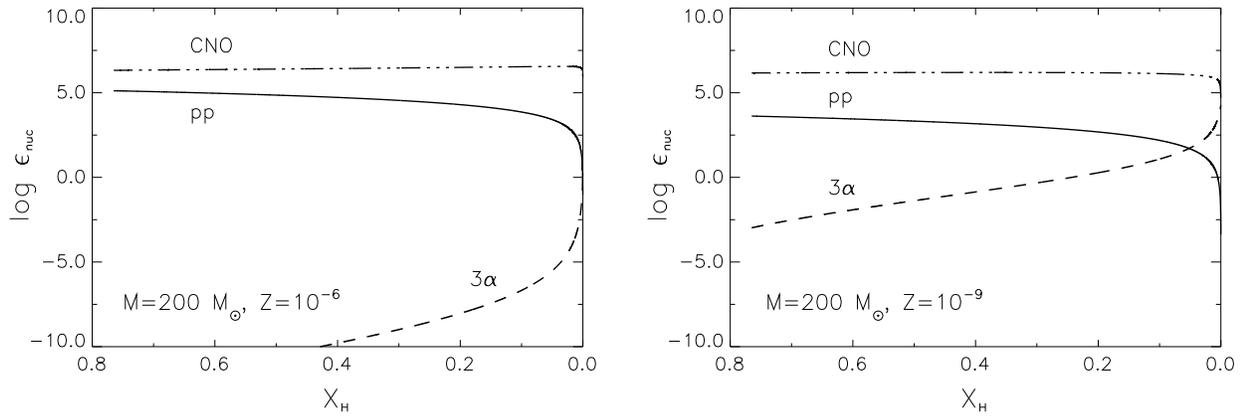}
%  \vspace*{174pt}
\caption{Main sequence nuclear energy generation rate as function of
the hydrogen mass fraction for $200 M_{\odot}$ galactic (left
panel), and pregalactic (right panel) Pop III stars, with
metallicity $Z=10^{-6}$ and $Z=10^{-9}$, respectively.}
\label{bar_fig24}
\end{center}
\end{figure*}

In Fig.~\ref{bar_fig23} the nuclear energy generation as function of
the temperature is shown, and Fig.~\ref{bar_fig24} shows this energy
generation during hydrogen burning for both galactic and pregalactic
Pop III stars.

\subsubsection*{\qquad c) Convective core size}

In their models, Marigo et al. (2001) found that if the convective
core grows, it eventually reaches the H-shell and engulfs some
hydrogen-rich material, which is rapidly burnt via the CNO-cycles.
This causes a flash that expands the core, so that central helium
burning weakens and the convective core recedes temporarily (in
mass). After the flash has occurred, the convective core starts
growing again.

In the present models, the same picture takes place for galactic Pop
III stars but it does not in the pregalactic case. In this case, the
central temperature at the transition of nuclear burnings is higher
than in the first one.

Hydrogen exhaustion in the centre of the star causes a progressive
contraction of the star and the shrinking of the convective core
which finally vanishes when $X_{\rm c} \sim 10^{-3}$, so that no
standard overall contraction phase is found \citep{Castellani1983}.
A maximum in the energy released by gravitation occurs during this
burning phase, when $X_{\rm c}=10^{-9}$, and $59\%$ of the energy
released by the star is supplied by contraction. Ignition of the
full triple-$\alpha$ cycle is only slightly delayed with respect to
hydrogen burning ignition in a shell, and once again the
gravitational contraction supplies energy to the star. Once the
$3\alpha$ chain has become fully efficient, a convective core is
again developed.

For the present work, pregalactic Pop III stars have higher
temperatures than galactic stars allowing a soon ignition of the
$3\alpha$ reactions. Then the transition from hydrogen to helium
burning is smooth because the convective core does not vanish.

\subsubsection*{\qquad d) Eddington luminosity}

Stellar models may become gravitationally unbound, i.e., the rate of
energy outflow at the surface exceeds the corresponding Eddington
luminosity.

In the present models, the Eddington $\Gamma$ factor is calculated
by estimating the ratio between the current stellar luminosity and
the Eddington luminosity. In both the hydrogen and helium burning
phases, the studied cases of $100$, $200$ and $300 M_{\odot}$
galactic and pregalactic Pop III stars evolve below the upper
Eddington luminosity limit. During hydrogen burning, the $\Gamma$
factor increases, reaches a maximum toward the end of this burning
phase and then decreases. The most massive stars evolve during the
helium burning phase closest to the Eddington limit.

\section{Conclusions}
\label{sec:conclusions}

For the present work a large number of ZAMS models have been
calculated showing the main physical variables as function of the
mass, and the metallicity. As the mass increases, ZAMS stars get
bigger, brighter, and less dense. Pop III stars get very hot and
compact as metallicity decreases.

When metal-free stars settles down on the main-sequence, they have
smaller radii, hotter cores, and higher effective temperatures than
metal-enriched stars. Like their counterparts, lower metallicity
stars become systematically cooler, larger and more luminous over
their hydrogen burning lifetimes. The most massive stars have
shorter lifetimes than less massive stars.

To study their properties and internal structure, stellar structure
models on the main sequence have been calculated, emphasizing the
case of metal-free stars which are compared with their Pop I and II
counterparts. Pop III stars are more centrally condensed, denser and
hotter. On the main sequence their different internal regions are
always below the upper Eddington luminosity limit.

The most important feature of the metal-free models is the high
temperature they maintain in their photosphere. These stars have a
high ionizing photon production rate. The ionization caused by these
stars is a direct result of their high effective temperatures. For
$M>100 M_{\odot}$, the studied metal-free stars have effective
temperatures $T_{\rm eff}\sim 10^{5}$ K. Consequently, they are very
efficient at producing photons capable of ionizing hydrogen and
helium. According to Bromm, Kudritzki and Loeb (2001), very massive
Pop III stars might have played a significant role in the IGM
reionization of hydrogen and helium at high redshifts.

Nuclear burning proceeds in a non-standard way. For stars with no
metals, the pp-chains and the CNO-cycles do not provide enough
energy to support the star, so when it reaches the main sequence
keeps contracting until the $3 \alpha$ process starts to operate,
halts the contraction and in a very small time produces enough CNO
elements so that the energy from the CNO-cycle is capable of
supporting the star, which expands and settles down to the main
sequence. The stars maintain core temperatures in excess of $T_{\rm
c}>10^{8}$ K which are high enough for the simultaneous occurrence
of the $pp$-chains, the CNO-cycles and helium burning via the $3
\alpha$ process.

For these stars, the radiative opacity in their envelopes is
reduced, and their core temperature is high, then the \textit{first
stars} are hotter and smaller than metal-enriched stars. On the
other hand, the opacity of stellar matter is reduced at low
metallicity, permitting steeper temperature gradients and more
compact configurations at the same mass.

Very massive ($M\gtrsim 100 M_{\odot}$) galactic and pregalactic Pop
III stars develop large convective cores with important helium core
masses $M_{\rm He}\sim 40 M_{\odot}$. These quantities are important
for explosion calculations \citep{Umeda2000} and synthetic
derivation of the SN yields \citep{Portinari1998}. Semiconvection
does not greatly affect the stellar structure during the
main-sequence phase. However, during the shell hydrogen burning and
helium burning phases, it plays a significant role, and the
evolutionary results depends on the adopted criterion and the input
physics of the models \citep{Chiosi1986}.

We have calculated evolutionary models for lower metallicity very
massive Pop III stars. The evolution of these stars is similar to
that of metal enriched stars but now the evolutionary tracks are
shifted to the left of the HR-diagram, i.e. the models are bluer
than their metal enriched counterparts. This is because massive Pop
III stars are hotter and very luminous. Pregalacic stars are denser
and hotter than galactic Pop III stars. However, during their
evolution these stars are more luminous than the first ones.

During the hydrogen burning phase, very massive galactic and
pregalactic Pop III stars evolve almost fully convective. They form
a large core with a different structure depending on metallicity.
Very massive stars are dominated by radiation pressure and electron
scattering. As a consequence of the high radiation pressure, the
convective core tends to be larger at higher masses and to expand as
the star evolves.

Because the stars radiate near the Eddington limit, radiation
pressure due to electron scattering opacity can become substantial.
For pregalactic Pop III stars, with the metallicity $Z=10^{-9}$
considered in this work, they develop a helium core $M_{\rm
He}=39.7$, $93.5$ and $142.4 M_{\odot}$, corresponding to initial
stellar masses of $100$, $200$ and $300 M_{\odot}$, respectively.
Then, according to Fryer et al. (2001) and Heger and Woosley (2002),
these stars will likely explode by pair-instability supernovae for
the $200$ and $300 M_{\odot}$ cases or collapse into black holes for
the $100 M_{\odot}$ case. However, $M<130 M_{\odot}$ stars could
explode like hypernovae.

The present evolutionary models have been calculated without mass
loss and with ro-rotation, in accordance with other existing points
of view. However, the uncertain role of mass loss can be viewed as
one of the major systematic uncertainties remaining in the study of
metal-free stars. In subsequent papers we will extend the present
study to mass losing and rotating models.
\bigskip

%%%%%%%%%%%%%%%%%%%%%%%% ACKNOWLEDGEMENTS %%%%%%%%%%%%%%%%%%%%%%%%%%%

{\small\emph{Acknowledgements:} This work has been partially
supported by the Mexican Consejo Nacional de Ciencia y
Tecnolog\'{\i}a (CONACyT), Project CB-2007-84133-F, and the German
Deutscher Akademischer Austauschdienst. DB also acknowledges CONACyT
for a Ph. D. grant.

%%%%%%%%%%%%%%%%%%%%%%%% BIBLIOGRAFY %%%%%%%%%%%%%%%%%%%%%%%%%%%%%%%%%%%

\nocite{*}
\bibliographystyle{spr-mp-nameyear-cnd}
\bibliography{biblio-u1}

%%%%%%%%%%%%%%%%%%%%%%%%%%%%%%%%%%%%%%%%%%%%%%%%%%%%%%%%%%%%%%%%%%%%

%-------------------------------------------------------------------

\label{lastpage}

\end{document}